\documentclass[fleqn,usenatbib]{mnras}

\usepackage{newtxtext,newtxmath}
\usepackage{graphicx}
\usepackage{float} 
\usepackage{stfloats}
\usepackage[T1]{fontenc}

\DeclareRobustCommand{\VAN}[3]{#2}
\let\VANthebibliography\thebibliography
\def\thebibliography{\DeclareRobustCommand{\VAN}[3]{##3}\VANthebibliography}

\usepackage{graphicx}	% Including figure files
\usepackage{amsmath}	% Advanced maths commands
\newcommand{\mstar}{$M_{*}$}
\newcommand{\mvir}{$M_{\rm{vir}}$}
\newcommand{\xspace}{ }

\usepackage{tabularx}
\usepackage{booktabs}
\usepackage{orcidlink}
\usepackage{CJKutf8}

\begin{document}
\begin{CJK*}{UTF8}{gbsn}
\title[Reaching for the Edge II]{Reaching for the Edge II: Stellar Halos out to Large Radii as a Tracer of Dark Matter Halo Mass}

\author[K. Leidig et al.]{Katya Leidig\orcidlink{0009-0003-8358-8320},$^{1}$\thanks{E-mail: kleidig@umd.edu} Benedikt Diemer\orcidlink{0000-0001-9568-7287},$^{1}$ Song Huang (黄崧)\orcidlink{0000-0003-1385-7591},$^{2}$ Shuo Xu (许朔)\orcidlink{0000-0002-4460-0409},$^{2}$
\newauthor
Conghao Zhou (周丛浩)\orcidlink{0000-0002-2897-6326},$^{3, 4}$ and
Alexie Leauthaud \orcidlink{0000-0002-3677-3617}$^{5}$
\\
$^{1}$Department of Astronomy, University of Maryland, College Park, MD 20742, USA \\
$^{2}$Department of Astronomy, Tsinghua University, Beijing 100084, China \\
$^{3}$Physics Department, University of California, Santa Cruz, CA 95064, USA \\
$^{4}$Santa Cruz Institute for Particle Physics, Santa Cruz, CA 95064, USA \\
$^{5}$Department of Astronomy and Astrophysics, UCO/Lick Observatory, University of California, 1156 High Street, Santa Cruz, CA 95064, USA\\
}

% These dates will be filled out by the publisher
% \date{Accepted XXX. Received YYY; in original form ZZZ}

% Enter the current year, for the copyright statements etc.
\pubyear{2025}

% Don't change these lines
\label{firstpage}
\pagerange{\pageref{firstpage}--\pageref{lastpage}}
\date{}
\maketitle

% Abstract of the paper
\begin{abstract}
The diffuse outskirts of brightest cluster galaxies (BCGs) encode valuable information about the assembly history and mass of their host dark matter halos. However, the low surface brightness of these stellar halos has historically made them difficult to observe. Recent deep imaging, particularly with Hyper Suprime-Cam (HSC), has shown that the stellar mass within relatively large projected annuli, such as within $50$ and $100$~kpc, is a promising proxy for halo mass. However, the optimal radial definition of this ``outskirt mass'' remains uncertain. We construct an HSC-like mock observing pipeline to measure the stellar mass density profiles of BCGs in the IllustrisTNG simulations. Our mock observations closely reproduce HSC profiles across six orders of magnitude in surface density. We then systematically measure stellar masses within different annuli and how tightly they are connected to halo mass. We find that stellar masses measured within simple apertures exhibit considerably more scatter in the stellar mass–halo mass relation than those measured within projected ellipsoidal annuli. We identify an optimal range of definitions, with inner radii between $\sim70 - 200$~kpc and outer radii between $\sim125 - 500$~kpc. We also introduce two halo mass-dependent Sérsic models for the average stellar halo profiles. We present a Sérsic-based fitting function that describes the profiles as a function of the halo mass, \mvir, with a median error of $54\%$. Adding the central stellar mass of the BCG as a second parameters slightly improves the accuracy to a median error of $39\%$. Together, these results provide fitting functions for BCG stellar halos that can be applied to future wide-field surveys to infer halo masses from deep imaging data.
\end{abstract}
% Keywords

\section{Introduction}

Dark matter halos, born from overdense regions of the universe, are the fundamental sites of galaxy and cluster formation. Accurately measuring the mass and distribution of these halos is essential for refining cosmological models and understanding the composition, structure, and evolution of the universe \citep{evrard_biased_1989, wang_cluster_1998, diemand_distribution_2005, vikhlinin_chandra_2009, rozo_cosmological_2009, abbott.2020}. Yet, despite their central role, dark matter halos are not directly observable with radiation. A key challenge in astronomy is therefore to infer their masses and assembly histories using observable ``tracers.'' 

Over the past several decades, a wide range of halo mass tracers have been developed, including X-Ray observations of the intracluster medium \citep{vikhlinin.2006, reiprich.2009}, submillimeter measurements of the thermal Sunyaev-Zeldovich effect \citep{sunyaev.1972}, satellite galaxy kinematics \citep{mckay_dynamical_2002, brainerd_mass--light_2003, conroy_evolution_2007, more_satellite_2009, lange_maturing_2019}, weak gravitational lensing \citep{leauthaud_weak_2009, becker_accuracy_2011, von_der_linden_weighing_2014, applegate_weighing_2014, grandis_impact_2019, umetsu_clustergalaxy_2020}, and optical richness \citep{andreon_scaling_2010, murata_constraints_2018,aguena_wazp_2021, maturi_amico_2025}. Among these, weak lensing is the most direct probe of the dark matter distribution, though it is limited by signal-to-noise and often restricted to massive nearby halos  \citep[e.g.][]{hudson_cfhtlens_2015, mandelbaum_strong_2016}. Optical richness, or the number of galaxies bound to a cluster, is often defined as the amount of galaxies above a given luminosity threshold within a specified aperture. However, while widely used, richness suffers from both projection effects and large intrinsic scatter in its correlation with halo mass, which together introduce substantial uncertainties \citep{erickson_influence_2011, costanzi_modelling_2019, sunayama_impact_2020, wu_optical_2022, nyarko_nde_impact_2025}. 

Another promising tracer of dark matter halo mass is the stellar mass of a cluster's central galaxy, or brightest cluster galaxy (BCG). To first order, BCG growth is closely related to the assembly history of its host halo \citep{behroozi.2019}, giving rise to the stellar-halo mass relation (SHMR). This relation has been well established across a wide range of methods \citep{hoekstra.2007, leauthaud.2011, more.2011, behroozi.2013, coupon.2015, zu.2015, vanuitert.2016, shan.2017, tinker.2017, kravtsov.2018}. However, measuring the total stellar mass of massive BCGs is notoriously challenging. Their extended, low-surface-brightness outskirts make it difficult to accurately define their physical extent and thus calculate their total stellar mass \citep{bernardi.2013, bernardi.2014, bernardi.2017, kravtsov.2018, pillepich.2018a, huang.2018_indiv_meas}. Moreover, a single global scaling relation does not capture the more nuanced ways in which the internal structure of BCGs reflects the assembly history of their host halos. These challenges motivate the development of alternative stellar-mass–based tracers that can more directly probe halo growth and assembly processes.

One promising avenue is to consider the stellar outskirts of the BCG rather than its total stellar mass. In the two-phase formation model of massive galaxies \citep{oser_two_2010, rodriguez-gomez.2016}, the inner stellar regions form early on, dominated by in-situ stars created through rapid, centrally concentrated star formation \citep{dokkum.2008, damjanov.2009, toft.2014,dokkum.2015, wellons.2016}. Once star formation quenches \citep{hopkins.2008,johansson.2009,conroy.2015}, late-time growth is dominated by the accretion of satellite galaxies, which deposit stars at large radii through tidal stripping and dynamical friction \citep{dokkum.2008, bezanson.2009, huang.2013, patel.2013}. These ex-situ stars dominate beyond $\sim50$ to $100$ kpc, and their spatial distribution correlates with halo mass and merger history \citep{rodriguez-gomez.2016, remus_outer_2017, pillepich.2018a}. 

Recent observational evidence supports this picture of stellar mass outskirts as a promising halo mass tracer \citep{huang.2022, kwiecien_improving_2025}. Using the Hyper Suprime-Cam (HSC) \citep{aihara.2018} deep imaging at $0.2 < z < 0.5$ , \citet{huang.2022} demonstrated that the stellar mass measured between $50$ and $100$ kpc from the BCG center correlates more tightly with halo mass (from weak lensing) than the total stellar mass within $100$ kpc. This suggests that the stellar outskirts encode information about the host halo not captured by integrated stellar mass alone, and could serve as an independent probe with constraining power comparable to richness \citep{xhakaj_cluster_2024}. Moreover, \citet{zhou_relationship_2025} showed that projection effects have minimal impact on the correlation between outer stellar mass and halo mass, further reinforcing the robustness of this observable. Detailed HSC analyses also reveal significant structural diversity in BCG outskirts \citet{huang.2018_indiv_meas, huang.2020}, hinting at varied mass assembly histories even within relatively homogeneous cluster samples.

These diffuse BCG stellar halos have been probed at unprecedented depth with HSC, and forthcoming wide-field surveys promise even greater reach. Early results from Euclid suggest that stellar halos will be measurable at $z=0.7$ out to $500 \rm{kpc}$ \citep{ellien_euclid_2025}, while the Roman High Latitude Wide Area Survey will extend surface brightness limits to $\mu _V \approx 30 \rm{mag/arcsec}^{-2}$, and the Vera. C. Rubin Observatory's Legacy Survey of Space and Time (LSST) will map stellar outskirts for thousands of galaxy clusters across the sky with magnitudes $\mu \ge  30 \rm{mag} \ \rm{arcsec}^{-2}$ by year 10 \citep{englert_intracluster_2025}. These next-generation datasets will open a new window on the outskirts of BCGs as halo mass tracers.

However to interpret such datasets, we need to build an accurate understanding of the connection between stellar and dark matter halos. The first paper in this series,\textit{ Reaching for the Edge I} \citep{li.2022}, established the reliability of outskirts measurements across multiple deep imaging surveys --- the Hyper Suprime-Cam Subaru Strategic Program (HSC-SSP), the Dark Energy Camera Legacy Survey (DECaLS), the Sloan Digital Sky Survey (SDSS), and the Dragonfly Telephoto Array. They demonstrated, through source-injection tests, that careful sky subtraction enables accurate recovery of surface brightness profiles out to $100\text{–}150$ kpc, in statistical agreement with the true input profiles. 

Complementary results from large volume cosmological simulations, such as the IllustrisTNG Project  \citep[TNG]{genel.2014, vogelsberger.2014, sijacki.2015, nelson.2015,  marinacci.2018, nelson.2018, pillepich.2018, springel.2018, naiman.2018} also strengthen this picture. Using Illustris, TNG100, and TNG300, \citet{xu_outskirt_2025} showed that the outskirt stellar mass measured between $50\text{--}100$ kpc correlates more strongly with halo mass than either total aperture stellar mass or total ex-situ stellar mass. Additionally \citet{ardila.2020} compared 2D BCG profiles out to $100$ kpc in HSC with Illustris, TNG100, and the \textsc{Ponos} simulation \citep{fiacconi_cold_2016, fiacconi_young_2017}, showing that simulations reproduce outer stellar masses reasonable well but display mismatches in the inner regions. Motivated by these results, complementary work by \citet{zhou_relationship_2025} showed that 2D stellar mass density profiles in TNG300 behave similarly or even better than 3D stellar mass profiles as dark matter halo mass proxies. Additionally, \citet{manuwal_inferring_2025} use TNG300 to establish a scaling relation between the dark matter surface density profiles and the stellar profiles out to $R_{200}$ for a sample of $40$ halos with $M_{\rm vir}> 10^{14.5}$

In this paper we explore the 2D profiles of a population of BCGs in the IllustrisTNG 50, 100, and 300 simulations. This serves as a natural extension of previous work in many ways. First, we create a mock HSC-like observing pipeline to measure the stellar mass profiles in TNG galaxies. This pipeline uses a selection of all stellar particles, including those along the line of sight, as in \citet{zhou_relationship_2025}, and an isophotal fitting routine as in \citet{ardila.2020}, \citet{xu_outskirt_2025}, and \citet{zhou_relationship_2025}. However, we extend upon these studies by adding both noise and a realistic satellite masking routine. Second, we model these profiles to larger radii than previously studied ($150 \, \rm{kpc}$ in \citet{ardila.2020} and \citet{xu_outskirt_2025} and $300$~kpc in \citet{zhou_relationship_2025}), reaching $500\,\rm{kpc}$. Third, we carry out a systematic investigation of different inner and outer radial apertures for defining outskirt stellar mass in order to analyze their halo mass constraining power. Finally, we introduce two halo mass dependent fitting functions to characterize our 2D profiles. Our goal is to predict the stellar mass definitions in the outskirts that provide the most robust tracers of total halo mass, enabling direct comparisons with current and upcoming observational surveys.

This paper is organized as follows. In Section~\ref{sec:data-and-methodology} we describe the simulations, mock observing pipeline, and construction of stellar mass density profiles. Section~\ref{sec:results} presents our main results: the impact of resolution and modeling choices on the SHMR and profiles, the comparison between mock-observed and HSC profiles, and the calibration of mass-dependent Sérsic models and optimal outskirts mass definitions. In Section~\ref{sec:conclusion} we discuss the implications of these results and summarize our conclusions.

\section{Data and Methodology}\label{sec:data-and-methodology}

In this section we describe the methodology for measuring and comparing stellar mass density profiles in both our observational and simulated datasets. We first describe the HSC Survey (Section~\ref{sec:HSC}), our observed sample of high mass galaxies (Section~\ref{sec:hsc-sample}), and the derivation of their stellar mass density profiles (Section\ref{sec:hsc-profiles-and-masses}). We then introduce the IllustrisTNG Simulation suite (Section~\ref{sec:illustris-tng-background}), detailing our two stellar particle extraction methods and mock observing routine (Section~\ref{sec:mock-observing}).

\subsection{The HSC Survey}\label{sec:HSC}

The observational data in this paper originates from the Hyper Suprime Cam- Strategic Survey Project \citep[HSC-SSP]{aihara.2018, aihara.2018a, aihara.2019}\footnote{\url{https://hsc.mtk.nao.ac.jp/ssp/}}, a wide-field optical imaging survey using the $8.2$ m Subaru Telescope. Specifically, we use $\sim 137 \rm{deg}^2$ of optical images from the WIDE layer of the \texttt{S16A} internal data release. The HSC multiband images have a remarkable depth, able to measure surface brightness profiles of massive galaxies down to $> 28 \ \rm{mag} \ \rm{arcsec}^{-2}$ in the $i$- band ($\sim 3-4$ mag deeper than SDSS). Combined with the survey's superb median seeing in $i-$band ($\sim 0.58$ arcsec full-width half-maximum), wide field of view ($1.5^\circ$), and fine pixel resolution ($0.168$ arcsec), this makes HSC ideal for studying faint stellar halos.

We use radial stellar density profiles, which are products of images from the \texttt{hscPipe 4.0.2} pipeline - a derivative of the LSST pipeline developed for HSC \citep{juric_lsst_2017, axelrod.2010}. For a comprehensive understanding of the data reduction process, we refer the reader to \citet{bosch.2018} and for details on the photometric performance see \citet{huang.2018_photo_perform}. The redshifts for our galaxy sample are photometric redshifts obtained using the \texttt{frankenz}  \citep{speagle.2019} algorithm, with performance summarized in \citet{tanaka.2018}.  To avoid contamination from saturated stars, all galaxies are filtered through bright star masks as detailed in \citet{coupon.2017}. Further details about the HSC data can be found in \citet{huang.2018_enviro_size,huang.2018_indiv_meas,huang.2020}. 

\subsection{HSC Massive Galaxy Sample}\label{sec:hsc-sample}

From the \texttt{S16A} data, we select a sample of massive galaxies at redshift of  $0.3 < z < 0.5$. This redshift range allows us to resolve the inner light profile while also limiting the background level oversubtraction in the faint outskirts ($r \sim 100 \ \rm{kpc}$). Additionally, within this range we can ignore any redshift evolution of the galaxy stellar population. Our sample uses a cut of the \texttt{CModel}-based stellar mass, $M_{*,\rm{cmod}} \geq 10^{11.2} M_\odot$, where $M_{*,\rm{cmod}}$ is based on the $M_* / L_*$ estimated by five-band SED fitting using \texttt{iSEDfit} \citep{moustakas.2013}. 

Our final sample contains $16,968$ galaxies. All galaxies have a 1D profile measured in $i$-band out to $> 100$ kpc \citep{huang.2018_indiv_meas}. This sample is similar to the samples used in \cite{huang.2018_indiv_meas} and \cite{ardila.2020}.

\subsection{HSC Profiles and Masses}\label{sec:hsc-profiles-and-masses}

We use the same method for extracting stellar mass density profiles from $i$-band images as presented in previous work \citep{huang.2018_indiv_meas, huang.2018_enviro_size, ardila.2020, huang.2022}. Readers should refer to these papers for the full technical details. First, we apply an empirical background correction to the images and mask out any nearby contaminating objects, before applying iterative $3\sigma$ clipping. Using the \texttt{Ellipse} isophotal analysis function from \texttt{IRAF}, we extract the 1D surface brightness profiles. We calculate the median flux density value along concentric isophotes with varying semi-major axis values, all with fixed ellipticity. This makes our profiles robust against the many faint objects surrounding these massive galaxies \citep{ardila.2020}.  Our 1D profiles are stable above $\sim 28 \ \rm{mag} \ \rm{arcsec}^{-2}$, which corresponds to roughly $\sim 100 \ \rm{kpc}$ for our sample of galaxies. Additionally, the inner $\sim 5\textrm{-}6 \ \rm{kpc}$ of the profiles are smeared due to the 1" seeing ($\sim 6$ kpc at $z=0.5$). 

We then convert the $i$-band surface brightness profiles to surface stellar mass density profiles. To do so we use the average $i$-band $M_*/L_*$ ratio derived from the SED fitting after we apply corrections for galactic extinction and cosmological dimming. This value is dependent on choices of IMF, synthetic stellar population library, and star formation history model, but since our sample of low-$z$ massive galaxies is dominated  by old stellar populations, these choices do not majorly alter our results. \citet{huang.2022} showed that conclusions about the outskirt stellar masses of these galaxies remain the same even if we replaced the stellar mass estimates with $k$-corrected $i$-band luminosities. In this work we assume a radially constant $M_*/L_*$. Since low-$z$ massive galaxies have shallow but negative color gradients \citep[e.g.][]{huang.2018_indiv_meas,wang.2019,montes.2021}, using the average $M_*/L_*$ will lead to underestimates of \mstar~in the center of the galaxy and overestimates in the outskirts. However, given that \citet{huang.2018_indiv_meas} found no clear dependence of color gradients on \mstar, this systematic will not influence the conclusions of our work.

\subsection{Illustris TNG }\label{sec:illustris-tng-background}

We use the IllustrisTNG suite of cosmological magneto-hydrodynamical simulations \citep{marinacci.2018,naiman.2018,nelson.2018,pillepich.2018,pillepich.2018a,springel.2018}. IllustrisTNG was run using the moving-mesh code \textsc{AREPO} \citep{Springel.2010}. TNG analytically models sub-grid physical processes such as star formation, stellar winds, gas cooling, supernovae, and active galactic nuclei, all of which together give reasonable matches to properties in observed galaxies \citep{vogelsberger.2014, pillepich.2018}. The suite provides boxes of both baryonic and dark matter only simulations at three resolutions in three volumes. The box sizes refer to TNG50, TNG100, and TNG300 with side lengths of 35, 75, and 205 cMpc$/h$. Here we use the highest resolution versions of each box with $2 \times 2160^3$, $2 \times 1820^3$, and $2 \times 2500^3$ resolution elements respectively, with TNG50 having the highest resolution, 100 times that of TNG300. We therefore interpret the differences in galaxy properties between box sizes to be resolution effects. The simulations adopt the \citet{planck_collaboration_planck_2016} cosmology, with $\Omega_{\rm{m}} = 0.3089$, $\Omega_{\rm{b}} = 0.0486$, $h = 0.6774$ and $\sigma_{\rm{s}} = 0.8159$. 

Halo catalogs from TNG are constructed using the Friends of Friends (FOF) algorithm with a standard linking length of 0.2 times the mean inter-particle separation in the simulation \citep{Davis.1985}. In this work we use the term ``FOF group'' to refer to halos found by the FOF algorithm in TNG. These halos may contain multiple over dense substructures called ``subhaloes,'' which are identified via the \textsc{SUBFIND} algorithm \citep{Springel.2001}. The most bound subhalo within a larger FOF group hosts the central galaxy, while other bound subhalos host satellite galaxies. Galaxies are then found within subhalos. The galaxy found in the most massive subhalo is called the central galaxy, while all other galaxies in a FOF group are called satellite galaxies.

To match the HSC galaxy sample, we select all central galaxies in TNG50, TNG100, and TNG300 with total stellar masses  $M_{*}> 10^{11.2}M_{\odot}$ or halo mass $M_{\mathrm{vir}}>10^{12.9}$. Here \mvir \xspace refers the spherical tophat perturbation collapse definition from \citet{bryan.1998}. This results in a sample of 42, 303, and 5242 galaxies in TNG50, TNG100, and TNG300 respectively. 

\subsection{Mock Observing}\label{sec:mock-observing}

\begin{figure*}
    \centering
    \includegraphics[width=1.0\linewidth]{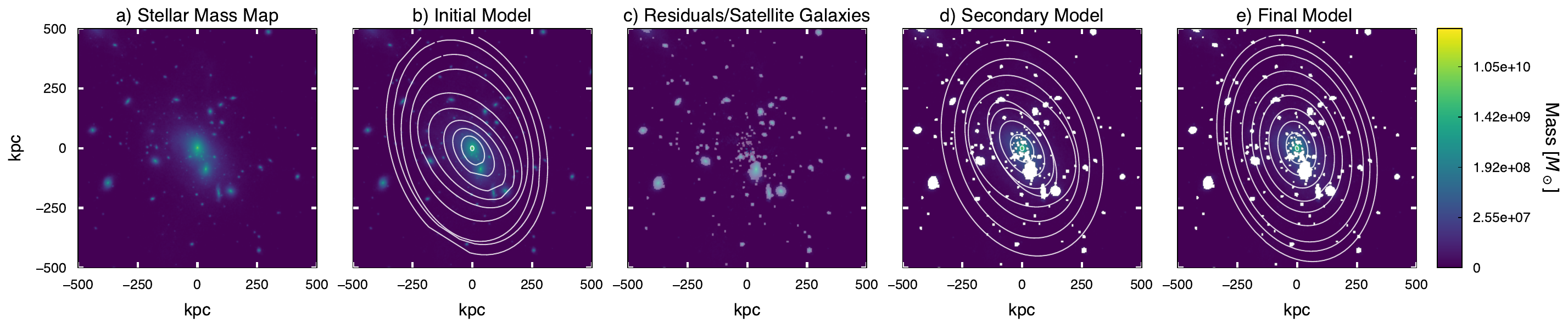}
    \caption{Example of our mock observing routine applied to a galaxy from TNG100. a) The projected stellar mass distribution, integrating all stellar particles over a $20$~Mpc line-of-sight depth centered on the galaxy's center of mass. b) The initial isophotal model of the central galaxy. c) The map of the residuals after subtracting the initial model from the total stellar mass distribution, with detected satellite galaxies highlighted in a lighter color. d) The secondary isophotal model following satellite subtraction. e) The final isophotal model of the central galaxy with constant ellipticity.} 
    \label{fig:example}
\end{figure*}

A central aim of this paper is to compare the stellar halos from the IllustrisTNG simulations with those observed in the HSC survey. To facilitate a fair and consistent comparison, we implement a mock observing routine modeled after the methodology used in the HSC data analysis. 

We begin by extracting a variety of scalar and vector quantities from our sample of IllustrisTNG galaxies. Some scalar properties—such as virial mass and radius are directly provided by \textsc{SUBFIND}. Others, including projected profiles and two-dimensional maps, are computed using the parallel data extraction framework \textsc{Hydrotools} \citep{diemer.2017a,diemer.2018,diemer.2019,tacchella.2019}. While these calculations for a galaxy are often performed using only the particles identified in that galaxy's \textsc{SUBFIND} defined subhalo within it's FOF group, this approach can underestimate properties at large radii where particles start to be lost to neighboring subhalos and even neighboring FOF groups. Because our analysis focuses on galaxy outskirts, we compute profiles using all particles within a given radius around a galaxy's center (``all-particle profiles'') and compare them to profiles using only the \textsc{SUBFIND}/FOF identified particles. For brevity we denote the latter as ``FOF-only'' profiles throughout this paper. This enables us to quantify any systematic bias introduced by the choice of particle selection. 

To enable efficient extraction of all particles around a center point regardless of FOF group or subhalo, we implemented a custom algorithm into the \textsc{Hydrotools} framework. Each simulation snapshot consists of multiple files containing particle information, with particles first organized in FOF groups and subsequently into subhalos, ordered from most to least massive. Fuzz particles, which are not bound to any halo, are stored after all the FOF groups in random order. Extracting particles purely on the basis of spatial coordinates is therefore impractical without opening every snapshot file individually. To address this, we constructed a lookup table that records the center of mass coordinates, radial extent, and file locations for each FOF group. We further divided the snapshot volume into spatial cubes and generated a secondary lookup table to record which files contain fuzz particles within each cube. Then for every galaxy of interest, the algorithm identifies the overlapping FOF groups and cubes within a user-specified extraction radius, reads in only corresponding particle files, and finally masks the subset of particles falling within the specified region. When a user-defined constant line-of-sight depth is requested, the extraction radius in increased accordingly to ensure all relevant particles are included. This ``all-particles'' routine was also used in \citet{zhou_relationship_2025}.

For each galaxy, we construct projected stellar mass maps out to $500$~kpc from the center of mass by projecting the stellar particles onto a $1000^2$ pixel grid, corresponding to a physical scale of $1.0 \rm{kpc/pixel}$. This matches the HSC observations at $z=0.4$. To account for projection effects, we create maps of in three orthogonal directions for every galaxy and treat these as independent realizations. This gives us total sample sizes of $126$, $909$, and $15726$ galaxy maps in TNG50, TN100, and TNG300 respectively. For each projection, we then create four separate stellar mass maps using different particle selections: (i) central galaxy particles only, (ii) satellite galaxy particles only, (iii) the combined central and satellite system (all FOF-bound stellar particles), and (iv) all stellar particles in the plane of the map with a line-of sight depth of $20,000~\rm{kpc}$. 

To replicate observational conditions, we take as input the stellar mass maps constructed using all particles around a central galaxy. We then process our maps using a routine that parallels the HSC analysis pipeline as laid out in \ref{sec:hsc-profiles-and-masses}.

We first add Gaussian noise with a standard deviation equal to the per-pixel $1 \sigma$ uncertainty, approximating the sky noise level in HSC imaging. We then take a first pass at finding the central galaxy geometry and center of mass using the \texttt{photutils} \texttt{Isophote} package \citep{bradley.2022}. This implements an iterative ellipse-fitting algorithm described by \citet{jedrzejewski.1987}. Although the galaxy center is, by construction, located at the map center, enforcing this assumption often introduces artifacts from the simulation such as negative gradients. Instead, we fit a series of isophotes with varying centers, ellipticities, and position angles to construct an initial model of the central galaxy (panel b of Figure~\ref{fig:example}). This isophotal modeling was also done by \citet{ardila.2020}, \citet{xu_outskirt_2025}, and \citet{zhou_relationship_2025}

We next implement a satellite masking routine. To do this, we subtract this initial model from the original map to create a residual map (panel c of Figure~\ref{fig:example}). In the residual, satellites and substructures are more easily identifiable. Using the \texttt{segmentation} tools in \texttt{photutils}, we compute a global background and noise estimate, convolve the data with a Gaussian kernel, and detect all sources above the background threshold. To separate overlapping sources, we apply watershed segmentation via the \texttt{deblend sources} routine. We take care to exclude the central region of the image from this process to avoid removing the BCG itself. Finally, we mask out the detected satellites (highlighted in a lighter color in panel c of Figure ~\ref{fig:example}) and remove them from the initial mass map. 

With the satellites masked, we repeat the isophotal fitting on the mass map with subtracted satellites. In this second pass, we also fix the isophote center to the mean of those obtained from the first pass, ensuring a stable central geometry. This secondary isophotal fitting yields a galaxy model with isophotes of varying ellipticity (panel d of Figure~\ref{fig:example}). For ease of calculations we construct a final model with constant ellipticity by averaging the geometric parameters of the fitted isophotes (panel e of Figure~\ref{fig:example}). The final mock-observed profiles are constructed by taking the average stellar mass density along isophotes of increasing semi-major axis lengths. We were able to successfully recover profiles for the majority of simulated galaxies ($99.2\%, 99.9\%, and 99.7\%$ for TNG50, TNG100, and TNG300 respectively), however inner regions that were too complicated to resolve.
    
% Results
\section{Results}\label{sec:results}
In this section we present our results and analysis. In section \ref{sec:shmr-in-3d-and-projected-space} we examine the impact of simulation resolution on the IllustrisTNG SHMRs and compare the mock-observed SHMRs to the HSC sample. Next, we examine the effects that particle selection, satellite subtraction, and annulus shape have on 2D stellar mass density, $\Sigma_*$, profiles in section \ref{sec:effects-on-sim-profs}. In section \ref{sec:comparison-of-project-profiles-in-sims-and-obs} we directly compare the mock-observed profiles to the HSC profiles in stellar mass bins. We next investigate an optimal stellar mass definition to use as a tracer of \mvir by comparing the SHMR slopes and scatter of various stellar mass definitions. Finally, we present a halo-mass dependent function to describe the extended profiles of our sample in section \ref{sec:profile-fitting}.
    
\subsection{Stellar Halo Mass Relations in 3D and Projected Space}\label{sec:shmr-in-3d-and-projected-space}

    \begin{figure*}
        \centering
        \includegraphics[width=1\textwidth]{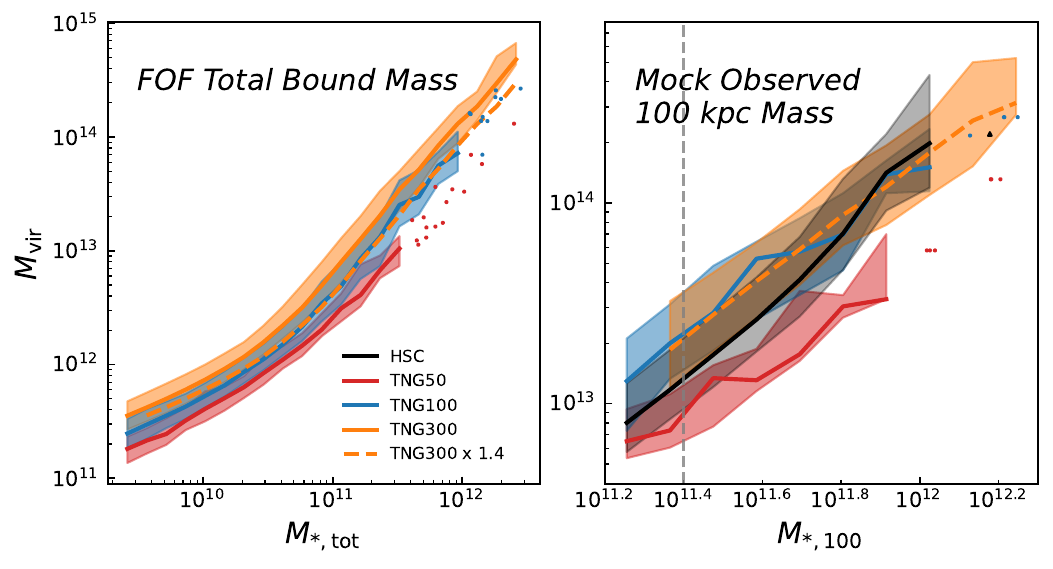}
        \caption{
        The Stellar Mass Halo Mass Relation (SHMR) for central galaxies in TNG50, TNG100, and TNG300 at z=0.4, showing the effect of simulation resolution on SHMRs. On the left are the SHMRs for central galaxies over a large mass range, while on the right are relationships for the sample of BCGs used in this study. \textbf{Left:} The median halo mass within bins of $0.15$ dex in stellar mass. Here the stellar mass is defined as the total 3D stellar mass bound to the central galaxy by \textsc{SUBFIND}. The shaded regions represent the $16^{\rm th}$ and $84^{\rm th}$ percentile scatter. the dots denote objects in bins too small to calculate a reliable median. The rescaled TNG300 stellar masses are shown by the dashed orange line. \textbf{Right:} The median halo mass as a function of stellar mass for the sample of BCGs used in this study. Here stellar mass is defined as the 2D stellar mass from mock-observed maps within a $100$ kpc semi-major axis. We compare the simulations to the observed HSC sample, which is plotted in black. At low masses, the TNG50 sample is the best match to the HSC sample, while at high masses, the rescaled TNG300 sample becomes a better match to the observations.}
        \label{fig:shmr}
    \end{figure*}

In Fig.~\ref{fig:shmr}, we present the $ z=0.4$ stellar-to-halo mass relations (SHMR) of central galaxies in TNG50, TNG100, and TNG300. As expected, the resolution differences between the simulations produce systematic offsets in their SHMRs, with TNG300 having the lowest stellar mass values. Following \citet{pillepich.2018a}, we increase the TNG300 stellar masses by a factor of 1.4 in order to match the TNG100 results in the halo mass range of $10^{12} M_{\odot}-10^{14} M_{\odot}$. We denote the rescaled values as rTNG300 (dashed orange line) and adopt them throughout the paper. We use TNG100 as a baseline instead of TNG300 because it has a higher mass and spatial resolution. At the other end of the resolution spectrum, consistent with \citet{engler.2021}, we find that TNG50 exhibits systemically higher central galaxy stellar masses than TNG100 (TNG300), by a factor of $\sim 1.5$ ($\sim 2$) at intermediate halo masses ($10^{11.5-12.5} M_{\odot}$), increasing to $\sim 2$ ($\sim 3$) at higher halo masses. Because the observed relation in HSC lies between TNG50 and TNG100, we do not apply any rescaling to TNG50.

\begin{figure*}
    \centering
    \includegraphics[width=1.0\linewidth]{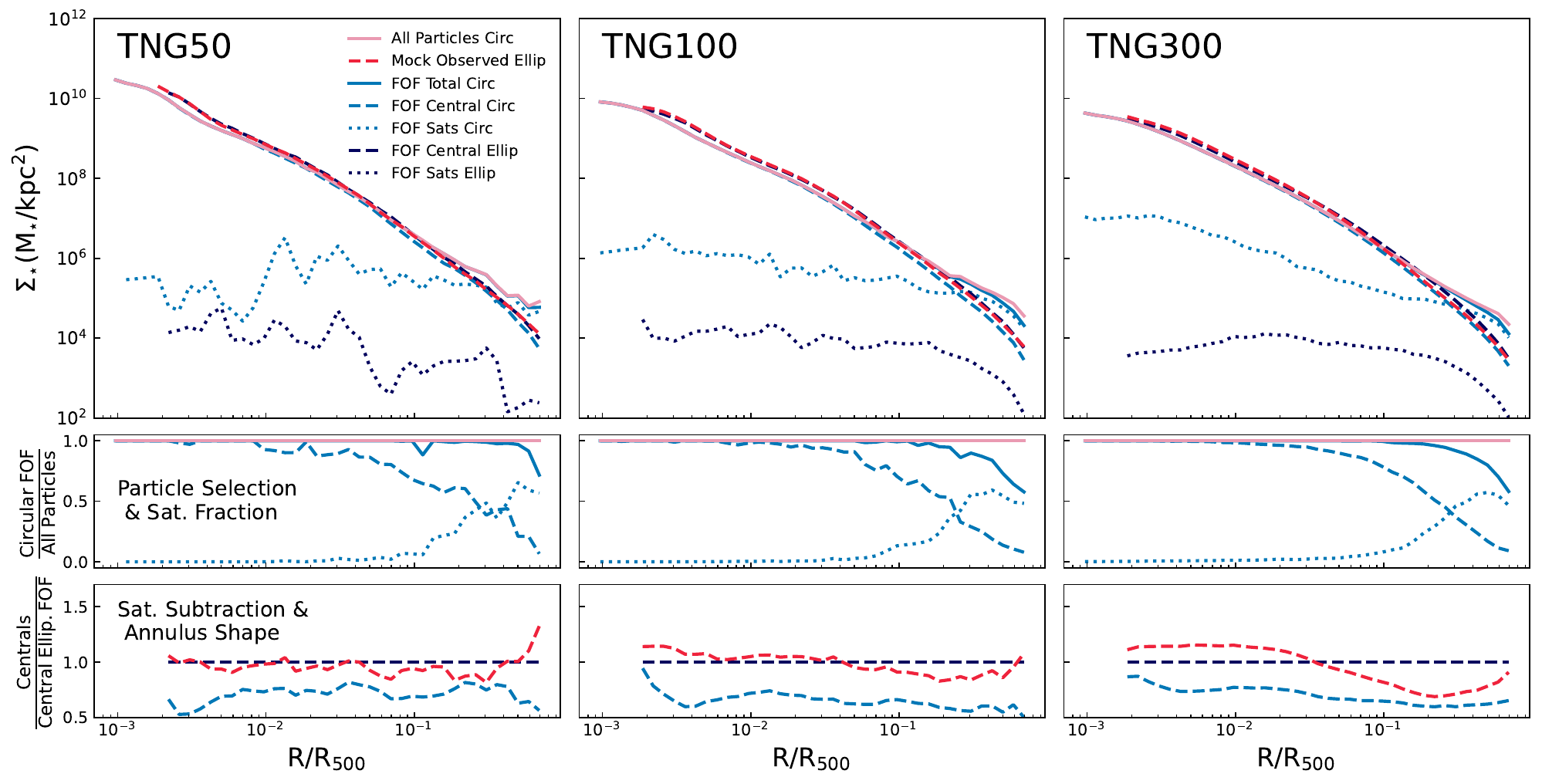}
    \caption{\textbf{Top:} Median stellar surface mass density profiles for halos in the mass range $13.2 \leq \log_{10}(M_{\rm vir}/M_\odot) \leq 13.5$, shown for TNG50 (left), TNG100 (middle), and TNG300 (right). Computed using different particles and geometries. \textbf{Middle:} Comparison of FOF (light blue) and all particle (pink) profiles calculated in circular annuli. Within the FOF sample, the central galaxy (dashed) dominated over satellites (dotted) out to ($\approx 0.01 R_{500}$). Beyond $\sim0.1,R_{500}$, the all-particle profiles (solid pink) exceed the FOF profiles (solid light blue), reflecting how the FOF algorithm assigns particles to subhalos, leading to a loss of material from the central profile at large radii. \textbf{Bottom:} Comparison of central galaxy profiles measured with circular (light blue) and elliptical (dark blue) annuli using FOF particles, and with our mock observing routine including satellite masking (red). Elliptical annuli yield systematically higher central densities (by nearly $50\%$ at the outskirts) compared to circular annuli. Satellite masking (red) raises the density at small radii and lowers it at large radii relative to the FOF-based subtraction (dark blue), highlighting differences in how subhalos are treated. Overall, these comparisons show that choices of particle selection, annulus geometry, and satellite treatment each produce systematic differences in $\Sigma_\ast$ profiles.}
    \label{fig:profs}
\end{figure*}

The left panel of Fig.~\ref{fig:shmr} shows SHMRs across the full mass range of each simulation. We include all central galaxies with $\log_{10}(M_{*,\rm{tot}}/M_\odot) > 9.34$, where $M_{*,\rm{tot}}$ is the 3D stellar mass bound to the galaxy's \textsc{SUBFIND} subhalo. This is a much broader mass range than that of our sample. We compute the median halo masses in bins of $0.15$ dex in $M_{*,\rm{tot}}$ containing at least seven galaxies; shaded regions indicate the $16^{\rm th}$ - $84^{\rm th}$ percentile scatter, while individual points mark galaxies in mass bins too sparse for a reliable median. Despite the resolution offsets, the three simulations exhibit similar shapes and scatters in their SHMR. Note that throughout this paper we plot the SHMR with stellar mass on the x-axis and halo mass on the y-axis, the opposite of the more common convention. We adopt this choice in order to systematically analyze the scatter in halo mass for a given definition of stellar mass. At lower stellar masses of $M_{*,\rm tot}\lesssim10^{10.5} M_\odot$ the scatter is $\sigma_{M_{\rm vir} | M_{*,\rm{tot}}}\sim0.25$ dex, increasing to $\sim0.4$ dex at higher stellar masses. The slope steepens from $\sim0.5$ at $M_{*,\rm tot}\approx 10^{9.5}M_\odot$ to $\sim1.4$ at $M_{*,\rm tot}\approx10^{11}M_\odot$, before flattening at the high-mass end. Here we plot the SHMR with halo mass on the independent axis.

The right panel of Figure~\ref{fig:shmr} compares simulated and observed SHMRs at the massive end, and the sample used in this paper. Before analyzing the detailed profile shapes, we want to establish that the overall mass normalization is consistent between simulations and the HSC data. To do this, we select BCGs with $\log_{10}(M_{*,\rm tot}/M_\odot) > 11.2$ and measure their projected stellar mass within a $100$ kpc semi-major axis ($M_{*,100}$) from the mock–observed maps, matching the stellar mass definition measured in the HSC data. The observed SHMR, derived from weak-lensing halo masses \citep{huang.2018_indiv_meas}, is shown in black. We compute the medians in bins of $0.11$ dex in $M_{*,100}$ with at least five galaxies. We find that the observed HSC relation lies between TNG50 and TNG100 at lower stellar masses and converges toward TNG100 at higher masses. 

All the simulated and observed samples show a scatter of $\sigma_{M_{\rm{vir}}|M_{*,100}} \sim 0.4$ dex. The simulations maintain a nearly constant SHMR slope across the probed mass range, whereas the HSC relation steepens at intermediate stellar masses before flattening again at the highest masses. Notably the observed SHMR transitions between matching TNG50 at lower stellar masses to aligning with TNG100 at higher stellar masses, indicating a reasonable but imperfect agreement. We note that our simulated sample may be incomplete at the lowest stellar masses considered ($M_{*,100} \lesssim 10^{11.4}M_\odot$), since galaxies were originally selected using $M_{*,\rm tot}$ rather than $M_{*,100}$. This however does not affect our conclusion at higher masses, which are the focus of this study.  

\subsection{Satellite Galaxy Subtraction and Ellipsoidal Fitting}\label{sec:effects-on-sim-profs}

Quantifying the systematics of how stellar mass surface density profiles are constructed is especially important when comparing simulated data to observations, and in particular when focusing on the galaxy outskirts. In Figure~\ref{fig:profs} we investigate the impact of four choices: (i) stellar particle selection (FOF/\textsc{SUBFIND} assigned vs. all stellar particles), (ii) the relative contribution of central and satellite galaxies, (iii) isophote geometry (circular vs. elliptical annuli), and (iv) the method of satellite subtraction (\textsc{SUBFIND} bound subhalos vs mock observed masking). To do so, we plot the median projected stellar density profiles of central galaxies in a representative halo mass bin, $10^{13.2} < M_{\rm{vir}}/M_\odot < 10^{13.5}$, corresponding to $3966$, $225$, and $24$ central galaxies from TNG300, TNG100, and TNG50 respectively. The median profiles are binned radially in steps of $0.07$ dex in $R/R_{500}$. All the profiles are shown in the top row of Figure~\ref{fig:profs}, while differences between certain profiles are shown in the middle and bottom rows. 

As discussed in section \ref{sec:data-and-methodology}, a common choice when defining the extent of a galaxy from IllustrisTNG is to use only the stellar particles bound to a \textsc{SUBFIND} subhalo within a FOF group. We investigate the effect that alternatively using all stellar particles in a given radius has on the profiles. In the middle row of Figure~\ref{fig:profs} we compare ``FOF-only'' profiles (light blue) with these ``all-particle'' profiles (pink), both measured in circular annuli. The two agree within $\approx~0.1 R_{500}$, but at larger radii the FOF-only profile contains up to a factor of two less mass, reflecting how the algorithm assigns star particles to neighboring subhalos or doesn't assign them to any subhalo at all. Decomposing the FOF profile into centrals (dashed light blue) and satellites (dotted light blue), as defined by \textsc{SUBFIND}, shows that satellites dominate beyond $\sim0.3R_{500}$, precisely where the FOF-only and all-particle profiles diverge. This indicates that the difference in FOF-only mass originates from how the algorithm treats satellites. 

The top row of panels in Figure~\ref{fig:profs} separates the profiles of central (dashed lines) and satellite (dotted lines) galaxies. We see that satellites contribute little inside $\sim0.1 R_{500}$ but rise sharply beyond this radius, becoming dominant at $\sim0.3R_{500}$. This is true for profiles in both elliptical (dark blue) and circular (light blue) annuli. This reinforces the point above: the outskirts of BCG profiles are highly sensitive to how satellites are assigned and whether all stellar particles are included.   

Another systematic is whether to compute profiles in circular or elliptical annuli. Since BCGs in at least the TNG300 simulation are majority non-spherical \citep{zhou_relationship_2025}, it is crucial to understand the difference between the methods. In the bottom row of Figure~\ref{fig:profs}, we compare the FOF-only central profiles in circular annuli (light blue) to those in elliptical annuli (dark blue), using the semi-major axis length as the radius. The elliptical annuli yield higher surface densities at nearly all radii, with the discrepancy growing at larger radii. This is expected: elliptical annuli follow the central galaxy’s shape and exclude more of the surrounding satellites, while circular annuli average over those regions. The difference is visible for both centrals and satellites (see satellite profiles in top row), confirming that geometry strongly influences the relative contributions. 
\begin{figure*}
    \centering
    \includegraphics[width=1\linewidth]{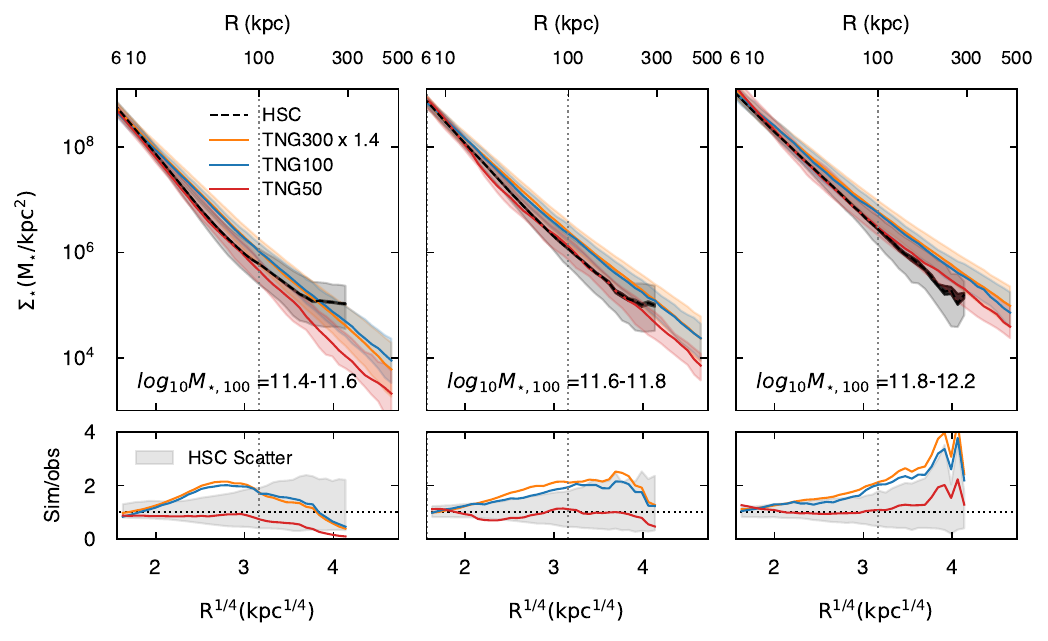}
    \caption{\textbf{Top}: Comparison of median stellar surface density profiles form HSC (black) and mock-observed central galaxies from TNG50, TNG100, and rescaled TNG300, shown in three stellar mass bins. Stellar mass is defined as the mass within a $100$ kpc semi-major axis of the galaxy center. Lighter shaded regions show the 16th and 84th percentile scatter. \textbf{Bottom} The simulated profiles compared to the observed profiles. The vertical line at 100 kpc shows the maximum extent where we are confident in the background subtraction of our HSC profiles \citep{huang.2018_photo_perform}. TNG50 matches the observations best in all mass bins.}
    \label{fig:compare}
\end{figure*}

Finally, we compare how satellite galaxies are treated in the FOF/\textsc{SUBFIND} catalog versus our mock observing routine, which identifies and masks satellites photometrically. In the bottom row of Fig.~\ref{fig:profs}, the mock-observed central profile (red, dashed) is contrasted with the FOF central profile measured in elliptical annuli (dark blue, dashed). At small radii, the mock-observed profile lies higher. This is likely due to a combination of two things. First, the satellite masking is intentionally disabled in the inner pixels to avoid overmasking—consistent with the HSC pipeline. And second, the mock-observed profiles include line-of-sight particles out to $20,000$ kpc. At intermediate radii ($0.03R_{500} \lesssim R \lesssim 0.6R_{500}$), however, the mock profile drops below the FOF profile, showing that the satellite masking removes a larger fraction of satellites than the FOF assignment. This trend works in the opposite direction from the all-particle vs. FOF-only comparison (pink vs. solid light blue in the top row), where including all stellar particles raises densities in the outskirts. These results slightly differ from those of \citet{ardila.2020}, who determined satellite masking method does not greatly impact the BCG's profile within $150\,\rm{kpc}$.

In TNG50, we see that at very large radii, $R > 0.4 R_{500}$, the mock-observed profile has higher densities than the FOF-only profile, whereas this behavior only appears at larger radii in TNG100 and TNG300. This shift may partly stem from TNG50's combination of higher resolution and smaller volume, which leads to a smaller clustering scale in TNG50 compared to the larger-volume runs where the SUBFIND subhalos are located closer to the central galaxy, meaning the ``extra'' stellar particles that the FOF/\textsc{SUBFIND} excludes show up at smaller radii. 

\subsection{Comparison of Projected Profiles in Simulations and Observations}\label{sec:comparison-of-project-profiles-in-sims-and-obs}
\begin{table}
    \centering
    \begin{tabular}{lccc}
    \toprule
    Sample & Bin 1 & Bin 2 & Bin 3 \\
    \midrule
          & \multicolumn{3}{c}{$\log_{10}(M_{*,100}/M_\odot)$} \\
          & 11.4–11.6 & 11.6–11.8 & 11.8–12.2 \\
    \midrule
    TNG50       & 32    & 21    & 11   \\
    TNG100      & 126   & 53    & 37   \\
    TNG300 ×1.4 & 2,386 & 1,142 & 608  \\
    HSC         & 4,962 & 1,348 & 229  \\
    \bottomrule
    \end{tabular}
    \caption{Number of central galaxies in each mock observed stellar mass bin for each simulation. Stellar mass $M_{*,100}$ is measured within a 100 kpc semi-major axis aperture.}
    \label{tab:mass_bins}
\end{table}

In Figure~\ref{fig:compare}, we compare the stellar surface density profiles of our mock-observed galaxies with those of HSC galaxies, binned by stellar mass. Each panel shows the median profiles of BCGs in bins of increasing $M_{*,100}$, defined as the stellar mass within an ellipsoidal isophote with a semi-major axis of $100$ kpc. The number of galaxies in each bin is listed in Table \ref{tab:mass_bins}.

The profiles are plotted in units of $r^{1/4}$ to highlight deviation from a de Vaucouleurs profile, traditionally used to describe elliptical galaxies. Solid lines denote the median profiles in 50 bins of equal width in units of $\rm{kpc}^{1/4}$. The lighter shaded regions show the scatter around the median, marked by the $16^{\rm th}$ and $84^{\rm th}$ percentiles. The reliability of these profiles is constrained at both small and large radii, owing to limitations in both the simulations and the HSC data. The profiles start at 6 kpc, below which the HSC data are unresolved due to seeing, and the simulated samples are affected by force softening. At large radii, HSC profiles are reliable out to $\sim 100$ kpc (dotted gray lines), beyond which background subtraction introduces systematic flattening. This flattening is less severe for the most massive galaxies. The bottom row of panels shows the linear difference between the simulated and observed stellar mass profiles. Solid lines again show median simulation profiles, while shaded regions denote the $1\sigma$ scatter of the HSC galaxies.

Overall, our mock-observed profiles agree well with HSC. In the two lower-mass bins, all simulations reproduce the observed profiles within a factor of two across six orders of magnitude in $\Sigma_*$; in the highest-mass bin, only TNG50 maintains this agreement. Across bins, TNG50 shows slopes consistent with HSC and remains within the observed $1\sigma$ scatter, while TNG100 and rTNG300 yield systematically shallower profiles. Although TNG50 overpredicts stellar mass at the low-mass end of the SHMR (Fig. \ref{fig:shmr}), this excess is concentrated in the innermost regions and is less evident in the radial profiles. Notably, the shapes of our mock-observed profiles are close to a Sérsic profile, a point we explore in detail in Section \ref{sec:profile-fitting}.

\begin{figure*}
    \centering
    \includegraphics[width=1\linewidth]{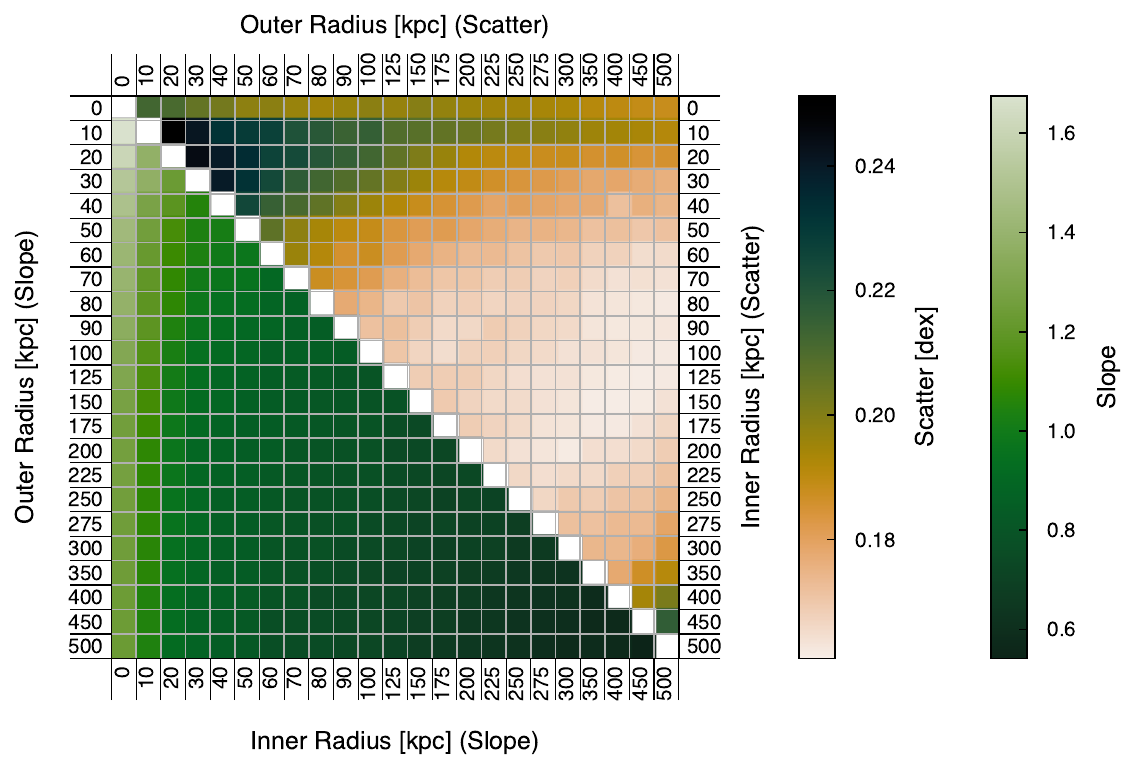}
    \caption{Slopes and scatters of the SHMR in TNG100 for various stellar mass definitions, showing which definitions are the best tracers of halo mass. The bottom-left triangle shows the SHMR slopes, with aperture mass definitions (which include the inner most core) being found in the left most column. The top-right triangle shows the scatter in $M_{\rm{vir}}$, with aperture mass definitions along the top row. We find the stellar masses with the smallest scatter in halo mass correspond to inner radii of 
    $\sim70 - 200$ kpc and outer radii of $\sim125 - 500$ kpc. This indicates that stellar mass measured in annuli towards the outskirts, rather than total aperture mass, provides the most robust tracer of halo mass.}
    \label{fig:grid}
\end{figure*}
Our findings are broadly consistent with the TNG–HSC comparison of \citet{ardila.2020}, but extend to larger radii ($500$ kpc vs. $200$ kpc) and present residuals in linear space. Finally, we confirm that profiles become progressively shallower with increasing stellar mass, consistent with \citet{pillepich.2014}, who showed that more massive halos host more extended stellar envelopes built up through satellite accretion and mergers.

\subsection{What Definition of Stellar Mass Best Traces Halo Mass?}\label{sec:what-definition-of-stellar-mass-best-tracks-halo-mass}

With the overall agreement between TNG and HSC established, we now use the simulations to test how well different stellar mass definitions trace halo mass. In particular, we extend beyond the $50 - 100~\rm{kpc}$ definition accessible to current data and explore how the slope and scatter of the SHMR depend on the chosen radial range of stellar mass around the BCG. Figure~\ref{fig:grid} shows the SHMR slope and scatter for a wide range of stellar mass definitions. 

\begin{figure*}
    \centering
    \includegraphics[width=1\linewidth]{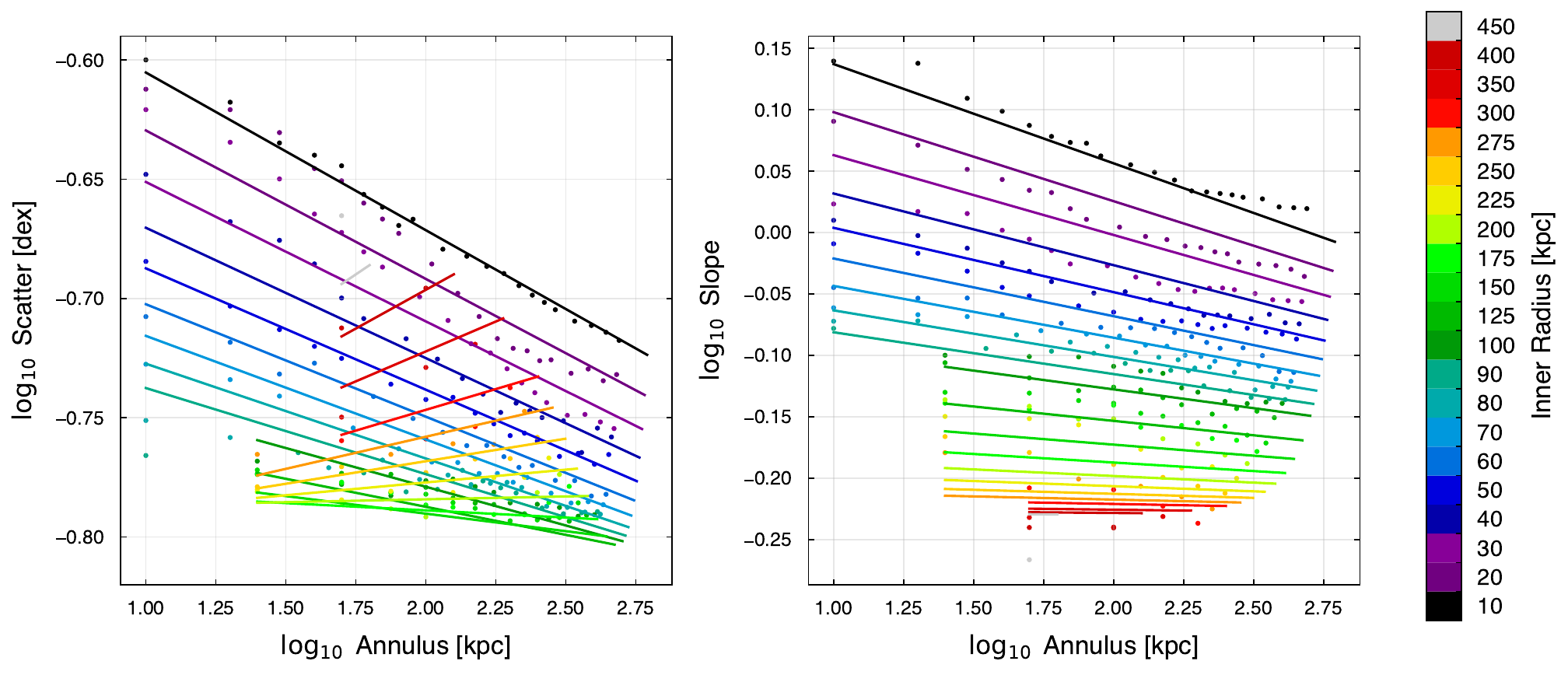}
    \caption{Trends in SHMR slope (left) and scatter (right) as they change with stellar mass definition, quantified by the mass definition's annulus width and inner radius. Points show measurements from TNG100, and solid curves show best-fit functions from equation ~\ref{eq: scatter} with best fit parameters found in Table \ref{tab:fitting_params}. The scatter is at a minimum at intermediate annular mass definitions, while the slope systematically decreases with inner radius.} 
    \label{fig:slopescatter}
\end{figure*}

We systematically define a range of stellar mass definitions, $M_{*,x-y}$ defined as the mass between an inner radius, $x$, and outer radius, $y$. We choose to use inner and outer radii in physical units of kpc as opposed to in units of effective radii (radius enclosing $50\%$ of total stellar mass). \citet{xu_outskirt_2025} found that effective radii based definitions performed the same or worse than physical definitions as halo mass proxies, showing comparable or larger scatter in their SHMR. While effective radii defined definitions sound promising for the purpose of self similar scaling, the uncertainties around their measurements seem to undermine their advantages.

Varying the stellar mass definition away from the one used in our initial sample selection introduces non-trivial completeness effects. To account for this, we construct a separate complete subsample for each definition of $M_{*,x-y}$ that only includes galaxies with $M_{*,x-y} > M_{*,x-y,L}$. We define $M_{*,x-y,L}$  by fitting a log-linear relation between $M_{*,x-y}$ and $M_{*,\rm tot}$ and then evaluating $M_{*,x-y}$ at $M_{*,\rm tot} = 10^{11.4}M_\odot$. We note that $10^{11.4}M_\odot$ is slightly higher than our original completeness threshold of $M_{*,\rm tot}= 10^{11.2}M_\odot$ and accounts for the fact that the mapping between $M_{*,x-y}$ and $M_{*,\rm tot}$ varies for different values of $x$ and $y$. However, a $0.2 $~dex adjustment is sufficient to include nearly all galaxies for most definitions. 

To compute the SHMR for each stellar mass definition, we use that definition's subsample and fit a log-linear relation to the median values of $M_{\rm vir}$ in $M_{*,x-y}$ bins. This method differs slightly from that used by \citet{xu_outskirt_2025}, who cut for all galaxies above the peak of the stellar mass distribution for each definition. This approach however, leaves itself susceptible to biases based on how the stellar mass is binned. While it does not impact the main results for stellar mass definitions with outer radii less than $150$ kpc, we choose not to implement this method for our broader range of stellar mass definitions.

The SHMR scatter is a measure of how well a given stellar mass definition constrains halo mass. In Figure~\ref{fig:grid}, the top-right triangle shows the scatter in $M_{\rm vir}$ for each $M_{*,x-y}$. Here the inner and outer radii defining the stellar mass definitions are denoted on the right and top axes respectively. The top-most row shows all definitions of $M_{*,x-y}$ where $x=0$, which include the central regions of the galaxy and are the most commonly used. We see that these aperture masses have smaller scatter than, for example, the mass definitions in the next row down where the inner radius, $x$, equals $10$~kpc, but a larger scatter than some of the more outskirt definitions. This result contrasts with \citet{huang.2022}, who found that in the observed HSC data, aperture masses produce larger scatter than annular definitions such as $M_{*,10-100}$. We confirm that the trends we find are intrinsic to the simulations, appearing in both TNG100 and TNG50 (though not TNG300, where inner structure is poorly resolved). We also note that our values for scatter are comparable to the values found by \citet{xu_outskirt_2025} for both aperture and outskirt definitions within $150$ kpc, and that similar to \citet{xu_outskirt_2025} we find that in general outskirt definitions have a smaller scatter than the aperture definitions.

The lowest scatter occurs for intermediate annuli, where the measurement excludes the central in-situ stellar core but remains wide enough to capture substantial stellar halo mass. These definitions typically have inner radii of $\sim70 - 200$ kpc and outer radii of $\sim125 - 500$ kpc. While our goal is not to select a single “best” definition, we note that stellar masses measured at these larger radii (e.g. $M_{*,125-350}$) trace halo mass more tightly than the $M_{*,50-100}$ definition currently accessible to HSC \citep{huang.2022}. 

To quantify the trend we see in the SHMR scatter across various stellar mass definitions, we fit a continuous function to the scatter as a function of the stellar mass definition taking the form of
\begin{equation}
    \log_{10}\sigma_{M_{\rm{vir}}} = \alpha(x) \log_{10}a + \beta(x)
    \label{eq: scatter}
\end{equation}
where $x$ and $a$ are the inner radius and annulus width ($y-x$) of the stellar mass definition respectively. $m$ and $b$ are functions of $x$ such that
\begin{subequations}
    \begin{align}
    \alpha(x) = \alpha_0e^{\alpha_1x}+\alpha_2 \\
    \beta(x) = \beta_0e^{\beta_1x}+\beta_2 \,.
    \end{align}
    \label{eq:slope_scat_components}
\end{subequations}
We choose to only fit to annular definitions of stellar mass, leaving out conventional aperture definitions ($x=0$). The best fit parameters for $\alpha_0, \ \alpha_1,\  \alpha_2, \ \beta_0, \ \beta_1 $ and $\beta_2$ are shown in Table \ref{tab:fitting_params}. The left panel of Figure~\ref{fig:slopescatter} shows this functional fit overlaid on the measured scatter values as a function of annulus, with the inner radius denoted by the color. The fitted function reproduces the trends we see in Figure~\ref{fig:grid}: the relatively high scatter for pure aperture measurements (where $x = 0$), the decrease in scatter as inner radius increases to intermediate values, and then the increase at very large inner radii where stellar density becomes faint and noisy. 

The lower left triangle of Figure~\ref{fig:grid} shows the slope of the SHMR relation for each definition of  $M_{*,x-y}$, where the inner radii and outer radii are denoted by the bottom and left axes respectively. Here, all the aperture definitions ($x=0$) are located in the left most column. There is a clear gradient in slope across all of our stellar mass definitions. We see the slopes become steadily shallower for definitions with larger inner radii, in line with the findings of \citet{xu_outskirt_2025}. Additionally for a given inner radius, the slope increases as outer radius (and hence annulus width) decreases. These trends follow such that as their ex-situ stellar fraction increases, their slope decreases. While the scatter of the SHMR is ultimately the indicator of how well a stellar mass definition traces halo mass mass, the slope can also offer some insight. A steeper slope means there's a large change in halo mass for a given change in stellar mass. While this large jump gives us a the leverage to distinguish between halo masses more easily, it also means that tracer is more susceptible to noise --- a small error in stellar mass would be a large difference in halo mass. On the other hand, a shallower slope yields less ability to distinguish between halo masses over a wide range of stellar masses. While there is no perfect slope for all purposes, the ideal stellar mass with intermediate values of inner and outer radii would exhibit a slope closer to unity. 

We quantify this trend in slope using the same equation we to describe the scatter (~\ref{eq: scatter}), this time fitting it to our slope values for all annular stellar mass definitions. The best fit parameters can be found in Table~\ref{tab:fitting_params}. The left panel of Figure~\ref{fig:slopescatter} shows the fit to the slopes as a function of annulus, again with the inner radii denoted by the colorbar. We see the dominate trend decreasing slope with increasing inner radius, and a secondary trend of slope decreasing slightly with increase annulus. 

Here we show only the SHMR slopes and scatters from TNG100, however we confirm that we see the same overarching patterns in TNG50 and TNG300, with modest differences. In TNG50, the increase in scatter for further annuli definitions begins at smaller inner radii, likely because the TNG50 sample has a lower median halo mass. By contrast, TNG300 shows a much smaller increase in scatter for the outermost definitions and does not exhibit the aperture definitions having smaller scatter than the annulus with an inner radius of $10$ kpc. This difference likely reflects TNG300's larger force resolution which causes scales $\leq 10~\rm{kpc}$ to be poorly resolved. Nonetheless the consistency of the overall patterns across TNG50, TNG100, and TNG300 demonstrates that our conclusions about optimal stellar mass definitions are robust against simulation resolutions. 

\begin{figure}
    \centering
    \includegraphics[width=\columnwidth]{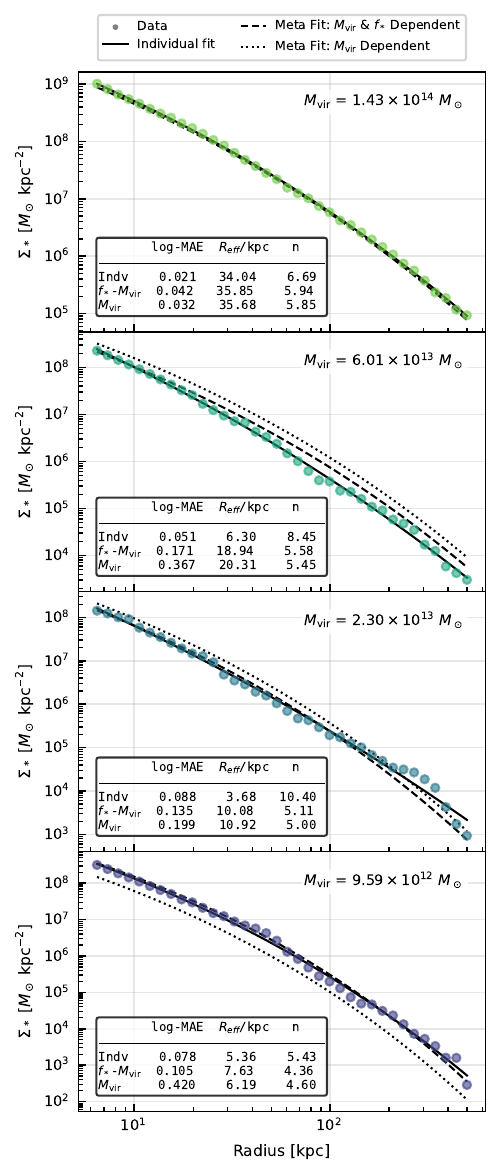}
    \caption{Stellar mass density profiles four random galaxies. The individual Sérsic fits are shown with solid black lines, the $M_{\rm vir}$ only dependent meta-model fit is shown by dotted lines, and the $M_{\rm vir}$-$f_*$ dependent meta-model fit is shown by the dashed lines. The values for $R_{\rm{e}}$ and $n$ are labeled for each fit as well as the fit's log Mean Absolute Error (log-MAE) (Equation~\ref{eq:mae}), where smaller error indicates a better fit. While the individual Sérsic fit has smaller error for all galaxies, the two halo mass-dependent equations fit pretty well, with the combined $M_{\rm vir}$-$f_*$ dependent model fitting slightly better for most galaxies than the $M_{\rm vir}$ only dependent model.}
    \label{fig:example_fits}
\end{figure}

\begin{table}
    \centering
    \begin{tabular}{lcc}
        \hline
        Parameter & Slope & Scatter \\
        \hline
        $\alpha_0$ & $-8.92 \times 10^{-2}$ & $-5.33 \times 10^{-1}$ \\
        $\alpha_1$ & $-1.11 \times 10^{-2}$ & $-7.29 \times 10^{-4}$ \\
        $\alpha_2$ & $-8.79 \times 10^{-4}$ & $4.63 \times 10^{-1}$ \\
        $\beta_0$ & $5.01 \times 10^{-1}$  & $3.24 \times 10^{-1}$ \\
        $\beta_1$ & $-1.12 \times 10^{-2}$ & $-1.01 \times 10^{-2}$ \\
        $\beta_2$ & $-2.31 \times 10^{-1}$ & $-8.32 \times 10^{-1}$ \\
        \bottomrule
    \end{tabular}
    \caption{Best-fitting parameters to describe the changes in SHMR slope and scatter across stellar mass definitions with varying inner radius and annulus values. See equation~\ref{eq:slope_scat_components}}.
    \label{tab:fitting_params}
\end{table}

\subsection{Mass-Dependent Profile Fit}
\label{sec:profile-fitting}

To characterize systematic trends in the stellar mass density profiles of our massive galaxy sample, we fit Sérsic profiles \citep{sersic_influence_1963} to the mock-observed 2D stellar mass profiles from TNG300. In order to avoid any interference from the gravitational softening, we exclude the inner $6~\rm{kpc}$ of the profiles ($\sim4$ times the TNG300 force softening length), but otherwise fit the entire range out to $500~\rm{kpc}$. The Sérsic model is described by 
\begin{equation}
    \Sigma(R) = \Sigma_{\rm e} \, 
    \exp \!\left\{ -\,b_{n} \left[ 
        \left( \frac{R}{R_{\rm e}} \right)^{1/n} - 1 
    \right] \right\} \,,
    \label{eq:sersic}
\end{equation}
where $\Sigma_{\rm{e}}$ is the surface density at the effective radius, $R_{\rm{e}}$, $n$ is the Sérsic index describing the profile shape, and $b_{n}$ is a constant chosen such that $R_{\rm{e}}$ encloses half of the total projected stellar mass.

\begin{figure}
    \includegraphics[width=\columnwidth]{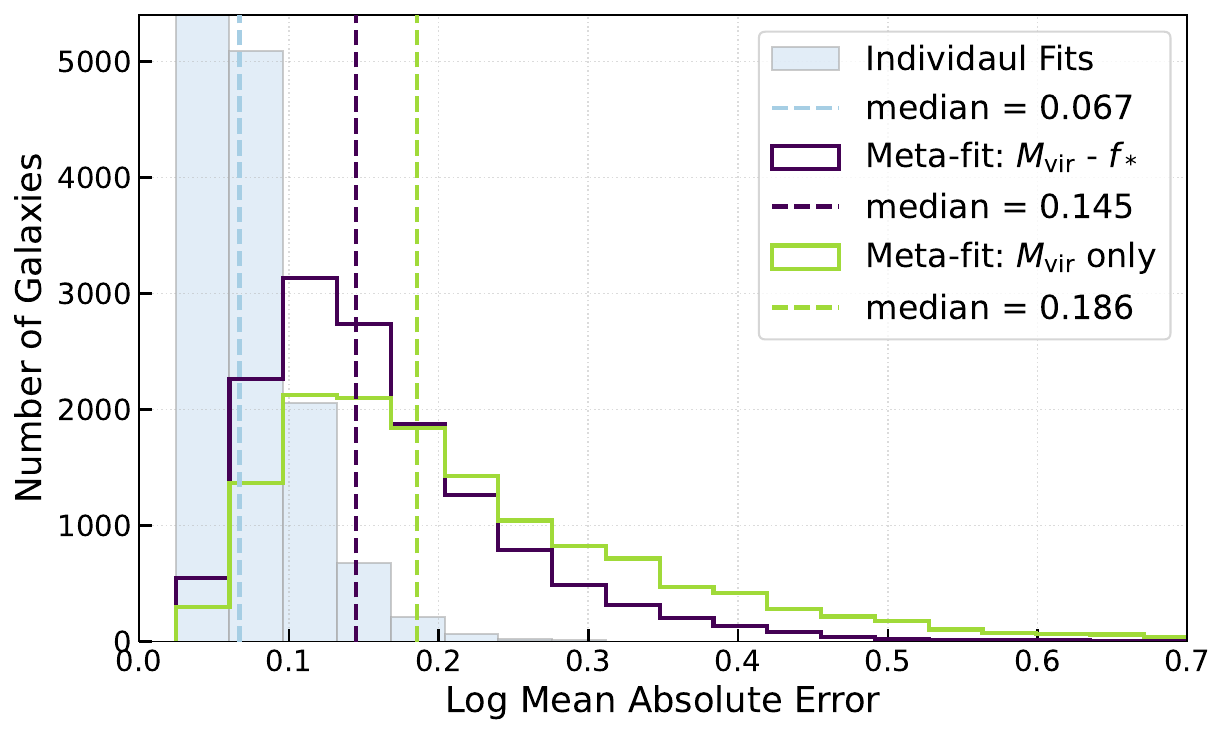}
    \caption{The distribution of logarithmic mean absolute error (log-MAE) for the individual Sérsic fits (blue) compared to the two meta models: $M_{\rm vir}$-only (green) and $M_{\rm vir}$-$f_*$ (purple). The median values for each distribution are denoted by dashed vertical lines. The median value of log-MAE for the individual Sérsic fits is $\sim 0.067$, corresponding to a $\sim 17\%$ difference from the mock observed profiles. The $M_{\rm vir}$-$f_*$ meta fit model yields a median log-MAE of $\approx 0.04$ less than the meta model with only $M_{\rm vir}$, equivalent to a change of $54\%$ difference to $39\%$ difference.}
    \label{fig:mae_hist}
\end{figure}

We find successful fits of the Sérsic model for $14129$ out of $15664$ mock-observed galaxy maps. The remaining fits either fail to converge or yield clearly nonphysical parameter values, so we exclude them from further analysis. Figure~\ref{fig:example_fits} shows the best-fit Sérsic model, denoted by solid black lines, for four randomly selected galaxies spanning a range of halo masses. The values for $R_{\rm{e}}$, $n$, and the log Mean Absolute Error (log-MAE) are printed for each galaxy. Here log-MAE is defined as
\begin{equation}
    \mathrm{log\text{-}MAE} = \frac{1}{N} \sum_{i=1}^{N} \, \left| \log_{10}(\Sigma^{\rm{model}}_{*,i}) - \log_{10}(\Sigma^{\rm{data}}_{*,i}) \right| \,,
    \label{eq:mae}
\end{equation}
where $N$ is the number of radial bins in each galaxy, and $\Sigma^{\rm{data}}_{*,i}$ and $\Sigma^{\rm{model}}_{*,i}$ are stellar mass densities for a given radial bin, $i$, from both the mock-observed profile and the modeled profile, respectively. In agreement with \citet{montenegro-taborda_photometric_2025}, we find that the Sérsic model is a relatively good fit for our profiles out to $500$~kpc. Galaxies with successful fits and $M_{\rm{vir}} > 10^{12.9} M_{\odot}$ ($13960$ galaxies) have a median value of log-MAE of $0.067$, which corresponds to a $\sim17\%$ difference from the mock-observed profiles. The distribution of log-MAE values for the individual Sérsic fits is shown in Figure~\ref{fig:mae_hist} in the shaded light blue. 

\begin{figure*}
    \includegraphics[width=\textwidth]{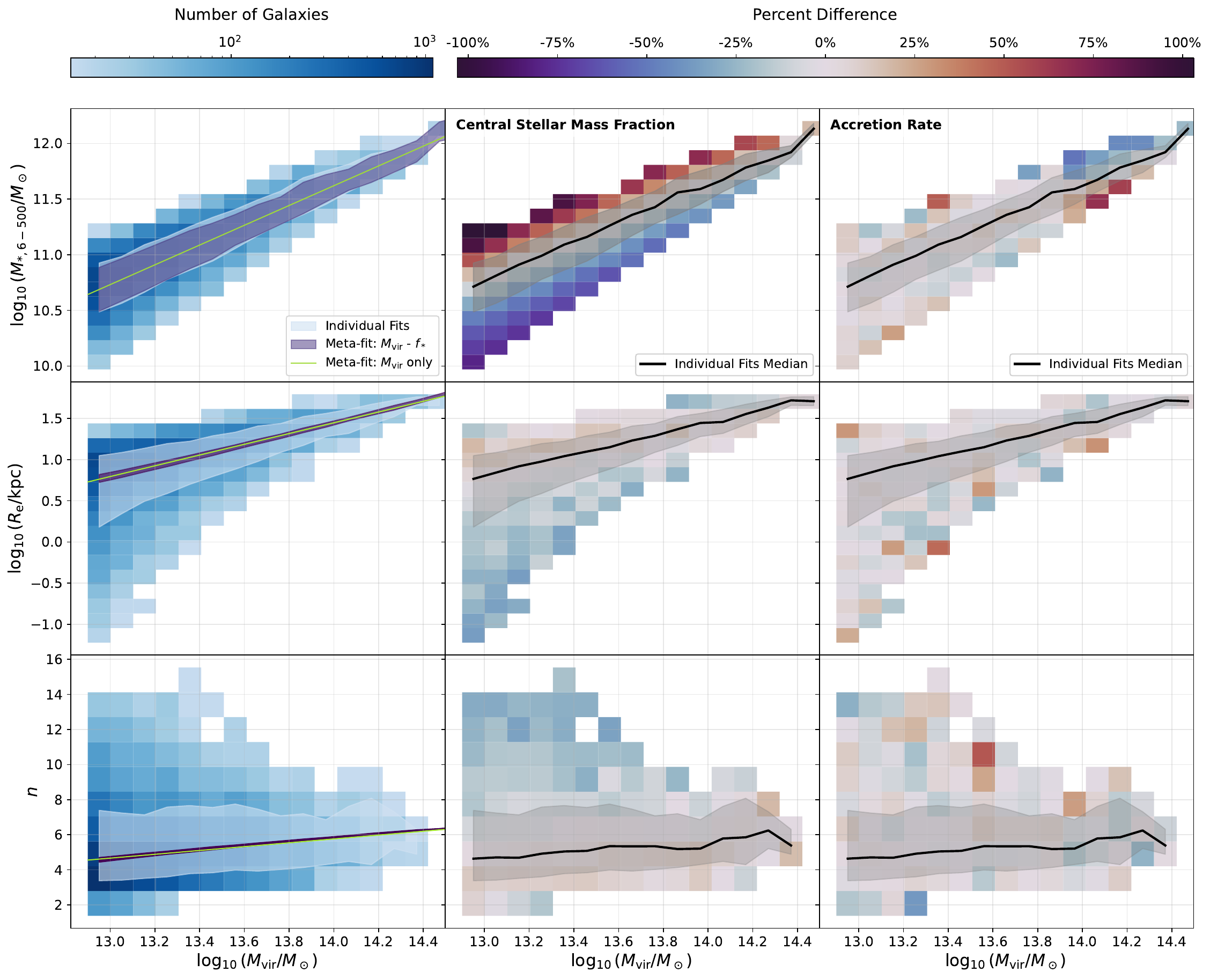}
    \caption{The integrated stellar mass $M_{*,6-500}$ (Row 1) derived from each galaxy's individual Sérsic fit, along with the corresponding Sérsic parameters $R_{\rm e}$ (Row 2) and $n$ (Row 3), are shown as functions of halo mass. \textbf{Column 1} shows the distributions colored by number density, with the $16^{\rm{th}}$ and $84^{\rm{th}}$ percentile scatter of individual fits indicated by the light blue shaded region. The results of the two best fit meta-models are overplotted: the $M_{\rm{vir}}$-only fit is shown by the green line, and the scatter in the $M_{\rm{vir}}$-$f_*$ fit shown by the purple shaded region. Both meta-models reproduce the median trends but fail to capture the scatter in $R_{\rm{e}}$ or $n$. \textbf{Column 2} colors the distributions by the percent deviation in central stellar mass fraction, $f_*$, from the mean at fixed halo mass. We find a strong correlation fo $M_{*,6-500}$ and weaker correlations in $R_{\rm{e}}$ or $n$. \textbf{Column 3} colors the distributions by deviations in accretion rate, $\Gamma_{\rm{dyn}}$, for which we see no significant trends.}
    \label{fig:3x3}
\end{figure*}

Given that the Sérsic model captures the overall shape of the stellar mass profiles remarkably well, we quantify how the resulting structural parameters scale with halo mass across our sample. Figure~\ref{fig:3x3} shows the distributions of $R_{\rm{e}}$, $n$, and the integrated stellar mass between $6$ and $500$ kpc, $M_{*, 6-500}$, as functions of halo mass. 

In the first column of Figure~\ref{fig:3x3}, each parameter's distribution is shown as a blue 2D number density histogram, with the $16^{\rm{th}}$ and $84^{\rm{th}}$ percentile scatter in each halo-mass bin indicated by the light blue shaded region. All three parameters show clear trends with halo mass: more massive halos have larger values of both $R_{\rm{e}}$ and $n$, indicating larger galaxies possess more extended stellar outskirts while still being more centrally concentrated (see \citet{graham_concise_2005} for a complete discussion and related mathematics). This joint increase of $R_{\rm e}$ and $n$ with galaxy size is well established for early-type galaxies \citep[e.g.][]{trujillo_estimation_2001}.

In addition to these first-order relations with halo mass, we also test for two potential second-order dependencies of stellar halo structure. The first is the central stellar mass fraction of the BCG, which we define here as 
\begin{equation}
    f_* = \frac{M_{*,30}}{M_{\rm{vir}}},
\end{equation}
where $M_{*,30}$ is the stellar mass within $30$~kpc. We choose $M_{*,30}$ as it is a common observable definition of BCG stellar mass \citep[e.g.][]{pillepich.2018}. The second column of Figure~\ref{fig:3x3} colors the 2D bins by the mean percent deviation in $f_*$ from the average value at fixed halo mass, thereby removing any underlying halo-mass dependence. As expected, $f_*$ shows a strong positive correlation with $M_{*,6-500}$. The relationships between $f_*$ and $R_e$ and $n$ are weaker but still present. At fixed halo mass, galaxies with higher $M_{*,30}$ (and therefore higher $f_*$) exhibit smaller values of $n$, and larger values of $R_{\rm{e}}$. These galaxies, which also have larger integrated stellar mass, $M_{*,6-500}$, are less concentrated with more extended and higher density outskirts.  

The other second-order dependency we investigate is halo mass accretion rate, which quantifies the recent growth of the dark matter halo. Rapidly accreting halos are expected to host BCG stellar halos that are physically smaller and exhibit shallower slopes at the outskirts \citep{deason_edge_2020}, enclosing a smaller fraction of the total halo mass \citep{ kimmig_intra-cluster_2025}. The accretion rate is measured over one dynamical time, $\Gamma_{\rm dyn} = \Delta \ln M_{\rm vir}/\Delta \ln a~|{t_{\rm dyn}}$, where $t_{\rm dyn}$ is the characteristic crossing time at the virial radius \citep{diemer_splashback_2017}. The third column of Figure~\ref{fig:3x3} colors the 2D bins by the mean percent deviation in $\Gamma_{\rm dyn}$ from the average value at fixed halo mass, again removing any underlying halo-mass dependence. We find no significant correlation between $\Gamma_{\rm dyn}$ and any of the Sérsic parameters or integrated stellar mass.

This is surprising because at fixed halo mass, halos with more recent accretion histories are expected to have less of their stellar mass incorporated into their BCG and intracluster light, and a larger fraction remaining in satellites, compared to earlier-forming halos \citep{purcell_shredded_2007, contini.2014, contreras-santos_characterising_2024, montenegro-taborda_stellar_2025}. At fixed halo mass, we may expect the integrated stellar mass, $M_{*, 6-500}$, to be larger for lower values of $\Gamma_{\rm{dyn}}$, since there has been more time for stars to be  disrupted from their satellite galaxies and accreted onto the BCG. Additionally, we may expect the halos with lower $\Gamma_{\rm{dyn}}$ to have stellar profiles that are less centrally concentrated and with larger extended outskirts. The absence of a clear trend warrants further investigation. 
 
Given the clear correlations of the Sérsic parameters with both halo mass and stellar mass fraction, we construct and fit two ``meta-models'' to quantify these dependencies. In the first model, the Sérsic parameters depend only on $M_{\rm{vir}}$. However, rather than allowing $\Sigma_e$ to vary directly with $M_{\rm{vir}}$, we choose to parameterize $M_{*,6-500}$ as a log-linear function of $M_{\rm{vir}}$, and derive $\Sigma_e$ such that 
\begin{equation}
    \Sigma_e = M_{*,6-500}/2 \pi (1-\epsilon)\int^{500}_6e^{-b_n(\frac{r}{R_e})^{\frac{1}{n}}-1}rdr,
    \label{eq: sigma_e}
\end{equation}
where $\epsilon$ is the ellipticity of the galaxy, and the integrated Sérsic profile reproduces the enclosed stellar mass. This approach ensures the fitted model corresponds to a physically consistent stellar-halo mass relation.

We parameterize this first model with $M_{*,6-500}$, $R_e$, and $n$ all as log or log-linear functions of $M_{\rm{vir}}$,
\begin{subequations}
    \begin{align}
    \log_{10}\!\left( \frac{M_{*,6-500}}{10^{14}\mathrm{M_\odot\,kpc^{-2}}} \right)
      &= m_{0} + m_{1}\,\log_{10}\!\left( \frac{M_{\mathrm{vir}}}{10^{14}\,\mathrm{M_\odot}} \right), \\
    \log_{10}\!\left( \frac{R_e}{\mathrm{kpc}} \right)
      &= r_{0} + r_{1}\,\log_{10}\!\left( \frac{M_{\mathrm{vir}}}{10^{14}\,\mathrm{M_\odot}} \right), \\
    n &= n_{0} + n_{1}\,\log_{10}\!\left( \frac{M_{\mathrm{vir}}}{10^{14}\,\mathrm{M_\odot}} \right),
    \end{align}
\label{eq: model-A}
\end{subequations}
with six free parameters $m_0, m_1, r_0, r_1, n_0,$ and $ n_1$. We fit this meta-model collectively to the mock-observed stellar mass density profiles of all galaxies in TNG300 with $M_{\rm vir} > 10^{12.9} \, M_\odot$ using a least-squares minimizer. The resulting best-fit parameters are listed in the first column of Table~\ref{tab:md_sersic}. 

Next, we fit a second model that varies the Sérsic parameters with both $M_{\rm{vir}}$, and the stellar mass fraction, $f_*$. We now parameterize $M_{*,6-500}$, $R_e$, and $n$ as
\begin{subequations}
    \begin{align}
    \log_{10}\!\left( \frac{M_{*,6-500}}{10^{14}\mathrm{M_\odot\,kpc^{-2}}} \right)
      &= m_{0} + m_{1}\,\log_{10}\!\left( \frac{M_{\mathrm{vir}}}{10^{14}\,\mathrm{M_\odot}} \right) \nonumber \\
      &\quad + m_2f_* +m_3 \log_{10}\!\left( \frac{M_{\mathrm{vir}}}{10^{14}\,\mathrm{M_\odot}} \right)f_*,\\
    \log_{10}\!\left( \frac{R_e}{\mathrm{kpc}} \right) &= r_{0} + r_{1}\,\log_{10}\!\left( \frac{M_{\mathrm{vir}}}{10^{14}\,\mathrm{M_\odot}} \right) +r_2f_*,\\
    n &= n_{0} + n_{1}\,\log_{10}\!\left( \frac{M_{\mathrm{vir}}}{10^{14}\,\mathrm{M_\odot}} \right) + n_2f_*,
    \end{align}
\label{eq: model-B}
\end{subequations}
with four new free parameters $m_2, m_3, r_2$, and $n_2$, for a total of ten free parameters. Fitting this model again to all galaxies in TNG300 with $M_{\rm vir} > 10^{12.9} \, M_\odot$ using a least-squares minimizer, we present the resulting best-fit parameters in the second column of Table~\ref{tab:md_sersic}. 

\begin{table}
    \centering
    \begin{tabular}{lcc}
        \toprule
        Parameter & $M_{\rm vir}$ only Model & $M_{\rm vir}$-$f_*$ Model\\
        \midrule
        $m_0$& $-2.377 \pm 0.004$&$-2.757 \pm 0.008$\\
        $m_1$& $0.885 \pm 0.005$&$1.111 \pm 0.010$\\
        $m_2$& --&$1.260 \pm 0.020$\\
        $m_3$& --&$0.537 \pm 0.020$\\
        $r_0$ & $1.451 \pm 0.010$&$1.408 \pm 0.013$\\
        $r_1$ & $0.646 \pm 0.013$&$0.717 \pm 0.19$\\
        $r_2$& --&$0.143 \pm 0.026$\\
        $n_0$ & $5.680 \pm 0.082$&$5.864 \pm 0.094$\\
        $n_1$ & $1.070 \pm 0.085$&$0.982 \pm 0.106$\\
        $n_2$ & --&$-0.355 \pm 0.124$\\
        \bottomrule
    \end{tabular}
    \caption{Best-fit parameters for the two mass-dependent Sérsic models. The $M_{\rm vir}$ only model is described by equation~\ref{eq: model-A} and the $M_{\rm vir}$-$f_*$ model is described by equation~\ref{eq: model-B}.}
    \label{tab:md_sersic}
\end{table}

In Figure~\ref{fig:example_fits} we compare the meta-model fits to the individual Sérsic fits for the four randomly selected galaxies. As expected, the individual fits lie closest to the mock-observed profiles, but both mass-dependent meta-models still capture much of the radial structure remarkably well. Figure~\ref{fig:mae_hist} quantifies this by showing the distribution of log-MAE values for all galaxies with $M_{\rm vir}>10^{12.9}$, comparing the individual fits (light blue), the $M_{\rm vir}$-only model (green) and the combined $M_{\rm vir}$–$f_*$ meta-model (purple). The $M_{\rm vir}$-only meta-model performs worse than the individual Sérsic fits, with a median log-MAE of $0.186\%$ (a $\sim54\%$ deviation). Including $f_*$ in the meta-model improves the agreement somewhat, reducing the median log-MAE to $0.144$, or roughly a $\sim39\%$ difference from the mock-observed profiles. 

To further assess the meta-models, we compare their predicted Sérsic parameters to those from the individual fits. In the first column of Figure~\ref{fig:3x3}, we show $M_{*,6-500}$, $R_{\rm e}$, and $n$ as functions of halo mass. Overlaid over the results of the individual Sérsic fits, the best-fit $M_{\rm vir}$-only meta-model is shown by a green line and the $M_{\rm vir}$-$f_*$ meta-model as a purple shaded band indicating the $16^{\rm th}$ and $84^{\rm th}$ percentile scatter. The $M_{\rm vir}$-only model treats $M_{*,6-500}$, $R_{\rm e}$, and $n$ as deterministic (log- or log-linear) functions of $M_{\rm vir}$, so it reproduces the median trends but introduces no scatter in these quantities, with any residual variation absorbed into $\Sigma_{\rm e}$. Adding the secondary dependence on $f_*$ allows the meta-model to generate scatter in $M_{*,6-500}$, $R_{\rm e}$, and $n$. The $M_{\rm vir}$-$f_*$ model does a relatively good job reproducing the observed scatter in $M_{*,6-500}$, but it still fails to capture almost any scatter in $R_{\rm e}$ and $n$, suggesting that further parameters or more flexible functional forms may be required to capture the full diversity of profile shapes. 

\begin{figure}
    \centering
    \includegraphics[width=\columnwidth]{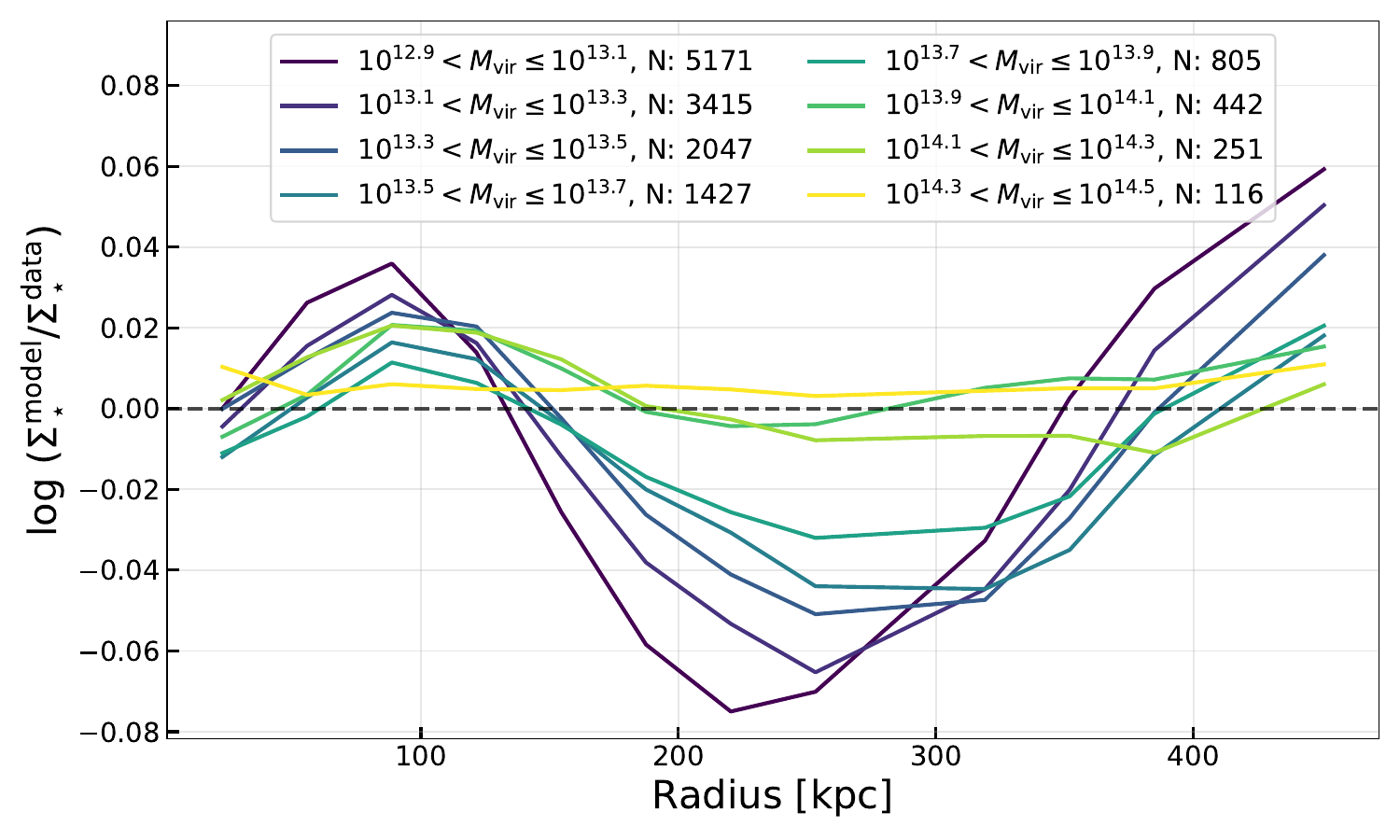}
    \caption{Mean residuals between mock-observed profiles and the $M_{\rm{vir}}$ only dependent meta-model across eight halo mass bins. The meta model shows systematic deviations, but overall is able to reproduce total stellar mass from the profile between $6$ and $500$ kpc.}
    \label{fig:residuals}
\end{figure}

Finally, it is useful to examine whether the meta-models introduce systematic errors as a function of radius. Figure ~\ref{fig:residuals} shows the residuals as a function of radius of stacked density profiles in eight halo mass bins for the $M_{\rm vir}$-only model. The mean binned residuals for the combined $M_{\rm vir}$–$f_{*}$ model are only marginally smaller, so we show only the $M_{\rm vir}$-dependent case here. Overall, the model reproduces the stellar mass profiles well, with residuals typically below $0.08$~dex. We find a systematic trend in halo mass: higher-mass bins show smaller residuals at all radii. The model also exhibits a characteristic radial pattern, overestimating the density between $\sim 50$–$150$~kpc, underestimating it between $\sim 150$–$350$~kpc, and again overestimating it in the extreme outskirts. Although detailed decompositions show that a single Sérsic profile often fails to capture the full light distribution of early-type galaxies \citep{huang_carnegie-irvine_2013, oh_testing_2017}, we find results consistent with the more recent work in \citet{montenegro-taborda_photometric_2025}, which demonstrated that a single Sérsic fit can provide a good description of the extended stellar halos of massive systems. Similarly, over the large radial range we consider here and for our high-mass halo sample, our parameterized single-Sérsic model offers a reasonable first-order description and successfully recovers the stellar mass between $6$ and $500$ kpc.

These results place our approach in the broader context of analytic descriptions of BCG and intracluster light (ICL) profiles. Observationally, \citet{montes.2021} model the 2D BCG+ICL light with a double-Sérsic decomposition. Going further, \citet{zhang_dark_2019} fit triple-Sérsic functions to azimuthally averaged surface-brightness profiles of over $300$ clusters, identifying components broadly associated with a compact core, bulge, and diffuse halo, a phenomenon previously noted by \citet{huang.2013}. On the modeling side, \citet{contini_brightest_2021} fit a combination of a Jaffé profile disk and a modified-NFW halo to 3D ICL profiles from a semi-analytic model. Many of these studies attempt to locate explicit transition radii between stellar components. In contrast, our method avoids hard boundaries and instead employs a single, mass-dependent model that captures the average structural trends.

\section{Conclusions}
\label{sec:conclusion}

We have combined deep observational data from HSC with realistic mock observations of the IllustrisTNG simulations to investigate how the extended stellar halos of BCGs trace their host halo mass. By modeling stellar mass density profiles out to $500 \, \mathrm{kpc}$, we have shown:  

\begin{enumerate}
    \item \textbf{Consistency between simulations and observations:} Mock-observed stellar mass density profiles from TNG, particularly TNG50, reproduce the shapes and amplitudes of HSC BCG profiles within a factor of two across the full radial range, validating both the simulation physics and our mock observing methodology.  

    \item \textbf{Systematic effects matter:} Choices in particle selection, isophotal geometry, and satellite subtraction each introduce measurable biases in simulated profiles, underscoring the importance of consistent pipelines when comparing to observations.  

    \item \textbf{Outskirts as optimal tracers:} Stellar mass measured in intermediate radial ranges with inner radii of $\sim70 - 200$ kpc and outer radii of $\sim125 - 500$~kpc correlates more tightly with halo mass than integrated aperture masses. We provide a fitting function for the SHMR slope and scatter as a function of the radial range and inner radius.

    \item \textbf{A mass-dependent Sérsic models:} We have introduced two Sérsic meta-models in which $M_{*, 6-500}$, $R_{\rm e}$, and $n$ scale either solely with $M_{\rm vir}$ or jointly with $M_{\rm vir}$ and the central stellar mass fraction $f_* = M_{*,30}/M_{\rm vir}$. Both frameworks successfully capture the systematic increase in profile extent and Sérsic index with halo mass, with the latter model additionally capturing the scatter in $M_{*, 6-500}$ and yields more accurate fits to the profiles. Together, these models can be used to forward-model observed profiles.
\end{enumerate}

The ability to model and interpret stellar halo outskirts at large radii will become increasingly critical as next-generation surveys (e.g., Euclid, Roman, Rubin) begin to deliver unprecedented wide and deep cluster profiles. \citet{dacunha_memoirs_2025} showed that the deep filters of Euclid, Roman, and Rubin will all be able to image stellar halos beyond $500~\rm{kpc}$ at $z =0.25$. By providing a framework to interpret observations, our results enable the use of BCG stellar halos as tracers of dark matter halos.

\section{Acknowledgments}
This material is based upon work supported by the National Science Foundation under Grant. No. AST-2206695. Additionally, this material is based upon work supported by the National Science Foundation Graduate Research Fellowship Program under Grant No. DGE 2236417. Any opinions, findings, and conclusions or recommendations expressed in this material are those of the author(s) and do not necessarily reflect the views of the National Science Foundation.

We are honored and grateful that this work is based upon data from the Hyper Suprime-Cam on the Subaru telescope on Maunakea, which has cultural, historical, and natural significance in Hawaii. 

The Hyper Suprime-Cam (HSC) collaboration includes the astronomical communities of Japan and Taiwan, and Princeton University. The HSC instrumentation and software were developed by National Astronomical Observatory of Japan (NAOJ), Kavli Institute for the Physics and Mathematics of the Universe (Kavli IPMU), University of Tokyo, High Energy Accelerator Research Organization (KEK), Academia Sinica Institute for Astronomy and Astrophysics in Taiwan (ASIAA), and Princeton University. Funding was contributed by the FIRST program from Japanese Cabinet Office, Ministry of Education, Culture, Sports, Science and Technology (MEXT), Japan Society for the Promotion of Science (JSPS), Japan Science and Technology Agency (JST), Toray Science Foundation, NAOJ, Kavli IPMU, KEK, ASIAA, and Princeton University.

This research extensively used the \textsc{python} packages \textsc{numpy} \citep{harris_array_2020}, \textsc{scipy} \citep{virtanen_scipy_2020}, \textsc{photutils} \citep{bradley.2022}, and \textsc{matplotlib} \citep{hunter_matplotlib_2007}.

% Bibliography
\bibliographystyle{mnras}
\bibliography{references_bibtex}

\bsp
\label{lastpage}

\end{CJK*}
\end{document}